# Potential Data Link Candidates for Civilian Unmanned Aircraft Systems: A Survey

Maede Zolanvari, *Student Member, IEEE*, Raj Jain, *Life Fellow, IEEE,* Tara Salman, *Student Member, IEEE*

***Abstract*— This survey studies the potential data link candidates for unmanned aircraft vehicles (UAVs). There has been tremendous growth in different applications of UAVs such as lifesaving and rescue missions, commercial use, recreations, etc. Unlike the traditional wireless communications, the data links for these systems do not have any general standardized framework yet to ensure safe co-existence of UAVs with other flying vehicles. This motivated us to provide a comprehensive survey of potential data link technologies available for UAVs. Our goal is to study the current trends and available candidates and carry out a comprehensive comparison among them. The contribution of this survey is to highlight the strength and weakness of the current data link options and their suitability to satisfy the UAVs communication requirements. Satellite links, cellular technologies, Wi-Fi and several similar wireless technologies are studied thoroughly in this paper. We also focus on several available promising standards that can be modified for these data links. Then, we discuss standard-related organizations that are working actively in the area of civilian unmanned systems. Finally, we bring up some future challenges in this area with several potential solutions to motivate further research work.**

## I. INTRODUCTION

Recently, there has been a spike in the number of applications of unmanned aircraft systems (UASs). These applications range from simple entertaining systems to sophisticated emergency medical services. Late in 2015, the Federal Aviation Administration (FAA) started requiring all the unmanned aircraft vehicles (UAVs) in the United States weighing in the range of 0.55 pounds to 55 pounds to be registered. Almost 800,000 registrations with FAA were recorded by the end of 2017. The FAA estimates that 1.9 million sales in 2016 would grow up to 4.3 million by 2020 for small recreational UAVs. Also, FAA foresees the growth in larger commercial UAV sales to increase from 0.6 million in 2016 to 2.7 million by 2020 [1]. Hence, the total anticipated growth is an increase from 2.5 million in 2016 to 7 million in 2020.

An obvious and yet the most crucial function of the UASs has been their ability to operate in emergency situations, where it is hazardous or expensive for human beings to get involved. This requires an effective real-time wireless communication link for the remote pilot to communicate with the UAS. Even in commercial or recreational applications, the UAV needs to be guided for accurate and safe navigation. Based on the specific application or the mission that UAS is supposed to be used for, the network and quality of service (QoS) requirements may vary.

A UAS consists of a UAV connected via a data link to a pilot on the ground. The pilot guides the UAV and may communicate with the air traffic control (ATC). By data link, we mean the connection link between the pilot and UAV, where the pilot may have direct access to the aircraft or indirect access through a network of data links such as cellular or satellite communications [2]. We will discuss these technologies later in this paper. The UAS architecture is shown in Fig. 1, where different dashed line types show different possible data links. The pilot may be in communication with the ATC to have a real-time situation awareness using a terrestrial link. In this paper, we do not discuss this terrestrial connection between the ATC and Pilot. Note that we use the term UAS for the entire system and UAV for the aircraft.

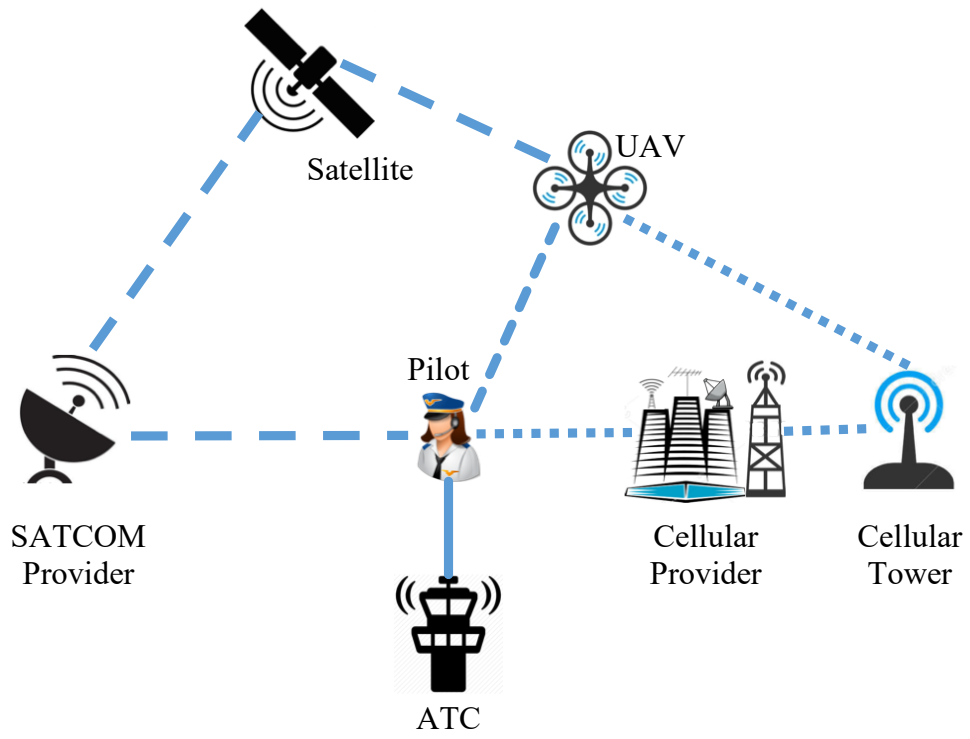

Fig. 1. UAS Architecture

### A. *Related Work*

Several aspects of the UAS's data link, such as security and safety concerns, network requirements, and channel modeling have been active areas of research and standardization. In this subsection, we briefly review some of the surveys that have been conducted. Due to the existence of these papers, these aspects are out of the scope of this paper.

#### 1) *Classification and Safety Concerns*

Regarding the UAS security and safety concerns, there are several survey papers available, including [3-8]. For example, a detailed discussion about safety considerations in using UASs while providing communication, navigation, and surveillance (CNS) services is studied in [5]. Different categories of classifications of UASs related to their safety requirements such as size, mission, and level of autonomy along with class demand forecasts are provided as well.

#### 2) Network Requirements

There are several survey papers discussing the requirements for UAS data link such as [2], and [9-14]. For example, the requirements for the UASs based on four different areas of missions, communication networks, communication data links, navigation, and surveillance are discussed in [2]. Several detailed discussions about the communication data link requirements such as range, velocity, latency, reliability, availability, integrity, security, and bit rate are provided. As another instance, a survey of UAVs communication requirements has been studied in [9]. These requirements are based on four categories of UAV applications (i.e., search and rescue, area coverage, delivery/transportation, and construction). Moreover, the important requirements and considerations regarding the network layer of UAV communications have been discussed in [14].

#### 3) Channel Modeling

Channel modeling must be studied for designing and evaluating proper UAV data links for different applications. There have been several research work and surveys on channel modeling for UAV communications, such as [15-18]. For instance, a comprehensive survey on channel modeling for UAV communications is provided in [15]. Different channel modeling approaches along with the associated challenges have been studied.

#### 4) Our Contribution

We have studied the existing survey papers on the aforementioned issues related to the UAV data links. In this survey, we focus on other aspects, specifically related to data link for civilian UAVs.

We provide a comprehensive discussion of several data link candidates and standards for unmanned systems, along with the standardization bodies active in this area. The aim is to explore existing proper protocols along with their pros and cons, compare the different candidates and highlight their possible utilization for UAV applications. Further, we discuss future challenges and potential solutions for a proper standard data link for UAVs.

The rest of this paper is organized as follows. In Section II, primary backgrounds related to the UAV data links are brought up, which are the basic required knowledge. In Section III, three general considerations and challenges related to data links for UAS are discussed. Section IV is about satellite communication (SATCOM). It introduces all the available satellite services providing communication data links suitable for UASs. Section V consists of information about cellular communication, along with technical details on different cellular generations. In Section VI, several data link standards that can be modified for use in UASs, current and future potential candidates, are studied thoroughly. In Section VII, standardization-related organizations active in this area are discussed. Future challenges along with some primitive solutions that need further research effort are highlighted in Section VIII. Finally, in Section IX, a summary of the paper is provided.

## II. BACKGROUND

This section discusses some background information that is important to understand the civilian UAV data links and their requirements. The rules and the restrictions of UAV flights in the airspace are first highlighted. Since in civilian applications, we mostly deal with small UAVs, the definitions and considerations of such systems are then provided. Following that, we highlight the importance of data link in these systems.

### A. Airspace Classes and Flight Rules

In 1990, the International Civil Aviation Organization (ICAO) defined an airspace classification, which is still used in the aviation industry. There are seven classes of airspace: A through G. Class G is the only uncontrolled airspace, which is defined for flights below 4,400 m above mean sea level (MSL). Flights in this class do not need any clearance. Hence, the operation of small UASs in this class is allowed without ATC permission. Operations in other airspace classes require ATC clearance to ensure that they will not conflict with other aircraft's flight paths. Also, all UASs within any airspace class require navigation accuracy for guidance and control.

The flight rules and the level of ATC interactions for each class are specified. Flight rules determine whether the aircraft can operate under visual flight rules (VFR), instrument flight rules (IFR), or special visual flight rules (SVFR).

In VFR regulations, the pilot must operate the flight in clear weather and have complete visions of where the aircraft is going. The minimum weather requirements for VFR are called visual meteorological conditions (VMC), and they are specific to each class. For example, distance from clouds is specified as clear of clouds (COC) for Class B.

IFR is another set of regulations for civilian aircraft's flights. For instance, in Class A airspace, all flights should be operated under IFR rules. SVFR is designed for the special case of operating under a set of VFRs for aviation; for example, operating the aircraft within a controlled zone in weather conditions below VMC. However, since there are more details related to this matter, we refer the readers to [19] for further information.

### B. Small UAVs

In a large portion of civilian UAV applications, small UAVs are deployed. Small UAVs, also known as "sUAVs," are the ones that bring the most restrictions and challenges when it comes to designing a proper data link. They are not allowed to fly in Class A airspace. As it was mentioned early, they mostly operate in Class G, and their maximum altitude is up to about 120 m (400 feet). Approximately 85% of the UAVs in the National Airspace System (NAS) are small, and most of the civilian UAVs are small. There are different definitions of what constitutes a small UAS, or sUAS. As defined in [20], FAA considers a UAV as "small" if its weight is less than or equal to 25 kg with a maximum mission radius of about 4.8 km (3 miles).

Since the sUAVs are relatively cheap, building a network of small UAVs to communicate as a mesh network with simple data links between them has been a popular approach in several

UAV applications. Currently, for small UAVs, wireless local area network (WLAN) links, including 802.11 (Wi-Fi), are popular, but their range is very short. The data transfer rate can be up to 274 Mbps, and the typical maximum range is between 30 m and 100 m based on the operational frequency.

The latest FAA's advisory circular on sUAVs regulations can be found in [21] that was provided in June 2016. Requirements and rules are fully explained. For instance, one of the requirements is that the sUAV must operate in the visual line of sight (VLOS) of the remote pilot during the entire operation.

### *C. Why Is Data Link Important in UAVs?*

UASs come with many challenges, including the need for reliable data links and autonomous controls. Although this is not a general assumption, in case of collision, the kinetic energy stored in a 25 kg UAV would instigate severe damage. That means even the small UAVs need a reliable data link to guarantee safe flights.

Many currently popular manned aviation applications use long-range satellite communications, which are expensive, and their large antennas are sometimes impossible to deploy on a small UAV. Employing any type of data links comes with specific advantages and disadvantages on the UAV's functionality regarding range, altitude, and payload.

It is also important to emphasize that the demands of the communication system in a UAS are highly dependent on the application and the mission that the system will be used for. Thus, the requirements of the data link will vary accordingly. Some of the popular civilian applications of UASs are shown in Fig. 2.

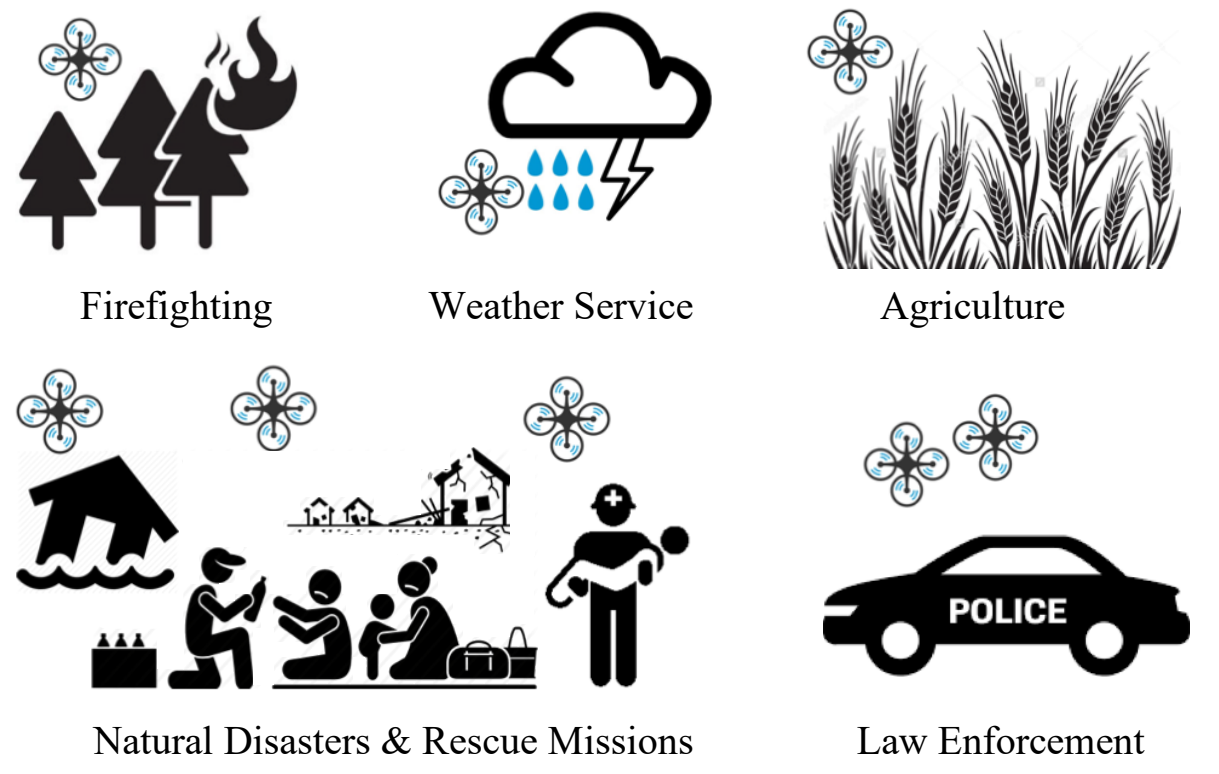

Fig. 2. Some civilian applications of UAS

Removing the onboard pilot from the aircraft in a UAS reduces the pilot awareness of the surroundings and aircraft condition. Therefore, the level of flight safety could decrease significantly. Even in manned aircraft vehicles, the automatic control modes need pilots to assist in providing the required level of performance and reliability.

Another issue related to UAVs' data link is with regards to their integration in the NAS. The performance differences between the UAS communication and other traffic types must be considered. This includes the differences in speed, range, and other flight aspects which complicate the ATC responsibilities to manage the co-existence of the manned and unmanned aircraft [22]. Hence, for ATC safety analysis, a calculated balance is needed for an unmanned aircraft. compared to the manned aircraft.

FAA Air Traffic Organization (ATO) is in-charge of providing the Certificates of Authorization or Waiver (COA) for commercial UASs (either small or large) to guarantee safe flights. However, there is no dominant communication standard or technology for UASs, so ensuring compatibility among different UAS platforms is difficult [23]. Moreover, there are no specific standards for UASs to use satellite or cellular communication as a data link. Defining a standard framework for beyond line of sight (BLOS) operations would boost the current interest for unmanned aviation even more than what has been already predicted. In this paper, we focus on the current trends for UAS data links with the hope to help standardization processes.

## III. UAV DATA LINK PRIMARY CONSIDERATIONS

As mentioned before, the data link is the most important part of the UAS. Without a resilient and reliable wireless communication link, it will not be safe for the UAV to operate and the mission outcomes would be unreliable and dangerous.

There are several primary considerations that should be included in designing the UAVs communication networks. These considerations highlight the importance of further investigations on general aspects to improve the scalability, safety, and QoS related to all the aerial vehicle systems.

In this section, we discuss five general aspects to be considered when choosing the best data link for civilian UASs; considerations on frequency spectrum and the available bandwidths for aviation; different considerations vital to payload and control links design; size, weight and power (SWaP) and resource allocation considerations related to design limitations of UAVs; considerations caused by its mobile nature and signal propagation; and the routing considerations.

### *A. Spectrum Considerations*

The most commonly used frequency bands in UASs' data links are K, Ku, X, C, S, and L bands. We discuss each of these briefly next.

K (18 to 27 GHz) band is a wide-range band that can carry a large amount of data, but it consumes a lot of power for transmission and is highly affected by environmental interferences. K, Ku (12 to 18 GHz), and Ka (27 to 40 GHz) bands have been mostly used for high-speed links and BLOS. Non-line of sight (NLOS) communication happens when the transmissions across a path between the receiver and the transmitter are partially or completely obstructed with a physical object, or simply is not in the line of sight. BLOS communication implies that the transmitter and the receiver are either too distant, usually as far as thousands of km, or too fully obscured, mostly because of the curvature of the Earth's surface, and the pilots should use cellular or satellite links.

The X band (8 to 12 GHz) is reserved for military usage, which is out of the scope of this paper [24]. C band (4 to 8 GHz) is the most popular band for the line of sight (LOS) data links. The weather conditions affect this band less than the other bands. However, due to its relatively short wavelengths and high frequency, the signal attenuation is relatively high, which leads to a considerable amount of power consumption. Frequency channel measurements for C band as the data link for UAVs are studied in [25]. Metrics such as received signal strength (RSS) and channel impulse responses (CIRs) are considered. S band (2 to 4 GHz) and L band (1 to 2 GHz) can provide communication links with data rates more than 500 kbps; their large wavelength signals can penetrate through the buildings transferring a large amount of data. Also, the transmitter requires less power for the same distance compared to high-frequency spectrums such as K band. S band radio propagation characteristics and measurements in UAS were studied in [26].

Recently, there has been a tremendous interest in moving to lower frequency bands for the civilian UAS data link. For wireless data transfers, 433 MHz and 868 MHz bands in many regions of the world and 915 MHz band in the United States are dedicated to send the telemetry data that can be utilized for the UAV communications [27]. These region-specific allocations were determined by International Telecommunication Union (ITU) to utilize the industrial, scientific and medical (ISM) band without requiring a license. Moving to the 915 MHz band in UASs is an efficient option for several civilian applications such as goods delivery in which the UAV must explore a long path. Further, frequency hopping spread-spectrum technique is generally implemented in this band. IEEE 802.15.4 is the basis of many protocols including ZigBee, that utilizes the 915 MHz frequency band.

This frequency band is also used in SiK radios for the autopilot drone products. These radios were first developed by 3D Robotics (3DR) on open source platforms. 3DR is a company active in manufacturing commercial drones using 915 MHz data links [28]. SiK radio links are capable of having up to 25% less bit error rate compared to the other currently popular UAV links, and the data latency is as low as 33 ms [29]. Another advantage of these radios is their small size and light weight that makes them suitable for sUAV applications.

A UAS communication model and simulation to analyze the link quality is presented in [30]. As expected, the UAVs operating in low frequency such as 915 MHz show better performance and suffer less from the free space loss. Different performance tests on a UAS with data links in 915 MHz, 2.4 GHz, and 5.8 GHz over an outdoor environment and a complex multipath environment have been studied in [31]. They provide a detailed comparison of these links.

Several test results on measuring and modeling 915 MHz channel for low-altitude (about 200 m above ground level) UAV have been presented in [32]. The capability of 915 MHz band in providing a high-capacity communication between the UAV and the remote pilot is empirically proven in that work. Table I summarizes the features of each frequency band.

TABLE I
FREQUENCY BANDS USED FOR UAV COMMUNICATIONS

| Frequency Band | Features |
|---|---|
| K, Ku, Ka bands | • Used for BLOS communication<br>• Mostly used in satellite-based data links<br>• A solution for congestion of lower frequency bands<br>• Cost-effective only for high-altitude UAVs |
| X band | • Reserved for the military |
| C band | • Standard Wi-Fi band<br>• Very popular for UAV communications<br>• Less affected by weather conditions<br>• Suitable for small UAVs and hobby drones |
| S band | • Popular for UAV communications<br>• Penetrates easily into buildings and structures |
| L band | • Penetrates easily into buildings and structures |
| Below 1GHz | • Suitable for low power and long-range UAV communications |

### *B. Payload and Control Links Considerations*

A UAS has two types of links, payloads links and command (or non-payload) link, which have different ~~features and~~ requirements. Even their downlinks and uplinks have different specifications. Therefore, each part of the data link should be studied separately. The command downlink for sending UAV information to the pilot is relatively simple and does not need a large bandwidth. For example, for a sUAS, it can be implemented by a few kbps general packet radio service (GPRS) modem [33]. The command uplink from the pilot to the UAV may need a larger data rate unless the UAV is autonomous. Both downlink and uplink need to be encrypted and robust against manipulation and jamming. On the contrary, the payload downlink is used to transmit the information such as status data, sensor data, and image data from UAV to the pilot and, therefore, requires much higher bandwidth than the command link. For instance, we may need 1 Mbps data rate for telemetry and video data, while 3 kbps may be enough for command and control download link [13].

For the command and control data links, reliability and robustness are the main design factors, whereas the data throughput may not need to be high [34]. Hence, the lower frequency band of the radio spectrum, where the robustness is high, would be a good choice. System integration in the lower frequencies is simpler, and the required transmission power is less. These bands suffer less from path loss; hence, the coverage range is higher. They are more compatible with the existing regulation for the aeronautical radio frequency (RF) spectrum. However, in the low-frequency parts of the radio spectrum, the bandwidth availability is limited; hence, the data rates are lower.

On the other hand, the payload data links need higher throughput and less resilient links. Thus, a higher frequency band is more suitable for these links. In summary, the type of links and mission requirements should be considered when choosing a frequency band for any UAS application.

### *C. SWaP and Resource Allocation Considerations*

Size, weight and power (SWaP) of the aircraft are other design considerations to help determine which data link should be used in the system. The data link technologies that provide high range and reliability without increasing the size, weight or power consumption of the system are always preferable. The SWaP considerations are more crucial for small UAVs compared to other UAV classes.

As an example, the limited onboard power in small UAVs lowers the payload capacity, and the useful operational range is limited by the power of the RF transmission [35]. To solve the problem of limited onboard power, a popular approach is to employ a large number of small, low-cost UAVs to cooperate and make a large-scale network. This design is referred to as "*multi-UAV network.*" It is especially useful in the case of natural disasters where the access to power may be very limited. Further, the approach makes the system robust against hardware failures and software malfunctions. It can also be self-sustaining by storing the data in the UAVs and sending it to the base station whenever a connection is established. Thus, the system is not dependent on having a real-time external communication link. This framework has been useful in mission-critical situations (e.g., natural disasters) [3, 36].

In a UAV-based network, the pilot should manage the critical responsibilities such as resource allocation. This task would solve several problems such as the transmission conflict among the UAVs (e.g., through polling techniques) and resource distribution among them [37]. Resource allocation is a joint optimization problem with goals such as minimizing the total transmission power and maximizing the throughput.

The on-demand flexibility and mobility of the UAVs come at the price of SWaP limitations. To manage these limitations, resource allocation techniques specifically for UAVs have been studied [38-40]. It is important to make sure that the optimal resource allocation would not sacrifice other performance metrics such as transmission rate, spectrum, optimal UAV's placement, user QoS, etc. However, as all the existing works mention that there is not enough research work covering all aspects of the resource allocation of UAV-based networks.

### *D. Signal Propagation Considerations*

Due to the mobile nature of the UASs, several challenges arise with the signal propagation, including the Doppler frequency shift, dynamic connectivity, antenna power, losses due to signal attenuation, multi-path fading, interference, and jamming.

The Doppler frequency shift is one of the important challenges in designing UAS data links. It is caused by the movement of the aircraft, which makes the received frequency at the ground station (GS) to differ from the sent frequency. The difference may be positive or negative depending on whether the aircraft is getting closer to or away from the GS. The performance of the data link is highly affected by the Doppler spread, which limits the UAV speed.

Furthermore, since UAVs are mobile, and their connectivity is dynamic, compared to the traditional wireless networks, their communication channel status changes more frequently. The degradation of signal to noise ratio (SNR) at the receiver is caused by the the propagation loss. The propagation loss due to the large distance between the UAV and the GS affects the throughput and error performance of the data link. All these effects are dependent on the communication channel properties. This issue highlight the vital role of a proper channel modeling for these systems.

On the other hand, in the future integrated airspace, data and ground platforms would need to be shared among the manned and unmanned aircraft. Due to this, UAVs might not have full access to the required bandwidth resources all the time [41]. As a result, compatibility and co-existence with manned aircraft must be considered [10, 11].

Vahidi and Saberinia [42] propose six different channel models for high-frequency UAV data communications. Different scenarios are built upon different types of UAVs and different environments in which they operate. The channel models are defined based on the Doppler properties and delay profiles. As a conclusion, Orthogonal Frequency-Division Multiplexing (OFDM) systems with a small number of subcarriers would provide the best performance in high-frequency UAV applications, due to large Doppler shifts.

Several diversity technologies are used in aviation to overcome signal degradations of data links. Frequency diversity, which is the most popular technique, uses multiple channels at different frequencies to transmit the same signal. In time diversity technique, the same signal is transmitted multiple times. And finally, in the path diversity technique, multiple antennas are employed on the receiver or transmitter side or both sides to send multiple copies of the same signal. The physical distance between these antennas must be considerable so that the signal would experience different channel properties on each path. However, employing diversity technologies in the system is a complex and costly technique [43]. The performance of path diversity by employing multiple UAVs using OFDM modulation of IEEE 802.11g protocol has been tested in [44]. It is shown that in practice the path diversity improves the UAS system's throughput significantly.

### *E. Routing*

Route-planning is a critical step in every applications of UAVs. The scheduled route must be low risk and low cost, while maintaining the mission goals. In a multi-UAV network, this consideration becomes even more complicated. For instance, these concepts must be studied carefully: avoiding any conflict among the UAVs, using minimum number of UAVs to cover the route and finish the specified task, time optimization regards to assigning a UAV to a specific part of a route while others cover the rest, etc.

While designing an efficient routing program in a multi-UAV network, it is important to solve the trajectory optimization. These is still a need to work on this issue, as there are only a few preliminary works that have been done [45]. There is a comprehensive survey in [46], focusing on routing protocols for UAVs. Routing in single UAV and multi-UAV networks are studied and compared. Performance of the popular existing routing protocols are reviewed in detail.

As mentioned before, small UAVs suffer the most from the resource constraints. In [47], the problem of optimal routing is studied, while the fuel constraints are taken into account. In the designed scenario, which is actually the case for most of the UAV application, the aircraft is supposed to visit several target points during its mission and refueling depots are also positioned in its way. Therefore, the goal of an optimal routing scheduling is to while the UAV fulfills its mission, it never runs out fuel.

There are several works researching on optimizing the user scheduling and UAV trajectory to increase the minimum average rate and throughput per user. The main objective of these papers is to minimize the number of required UAVs to cover a specific area with a multi-UAV network [48, 49]. Autonomous flying approaches in a network of multiple UAVs to optimally locate the UAVs in the network are described in [50]. In [51], the propagation loss and the interference caused by all UAVs in a multi-UAV network have been studied.

### F. *Summary*

Even though choosing the best data link for UAVs is mostly dependent on the application and mission, there are several general considerations that must be assessed. These considerations include frequency spectrum congestion, which limits our choice of the frequency band in which the UAV must operate; differences between the payload and control links; their design constraints based on the limited onboard power and weight; the UAV's mobile nature that complicates their radio design; and optimized routing.

In the following sections, we discuss various technologies that can be used in an unmanned system for better guidance and a higher level of safety for UAV flights.

## IV. SATCOM

Aerospace Industries Association (AIA) estimates that the integration of UAS into civil airspace would generate $89 billion over the next decade. Based on the statistics in their report, each year more than 1 million hours of unmanned flights occur in the United States. As their integration with other flying vehicles in the NAS gets standardized, this number would grow even more quickly [52]. Satellite Communication (SATCOM) helps pave the way for UAV integration into the national airspace. SATCOM is data transmission link through satellites in different orbits at different distances from the Earth.

In this section, we will highlight the advantages, disadvantages, the standardization process, and available services providing SATCOM data communication links for UAVs.

### A. *Advantages*

SATCOM allows the UAS command, control, and payload communications to go BLOS of the pilot, providing the maximum realizable coverage range. The case of BLOS in a UAS is shown in Fig. 3. As shown in this picture, the user can expand its sight coverage by removing the LOS communication link.

Some basic advantages of SATCOM are extreme mobility, strong reliability, and jam-resistant communication along with high data rate. These features would benefit UAVs with video communications or image sensors for mission-critical tasks. Low earth orbiting (LEO) satellites operating at 2,000 km height and geosynchronous earth orbiting (GEO) satellites operating at 35,000 km height are the two popular SATCOM technologies used for UAS so far. Providing over-the-horizon view in the UAV applications expands their functionality and helps these systems to be applied across all aviation applications.

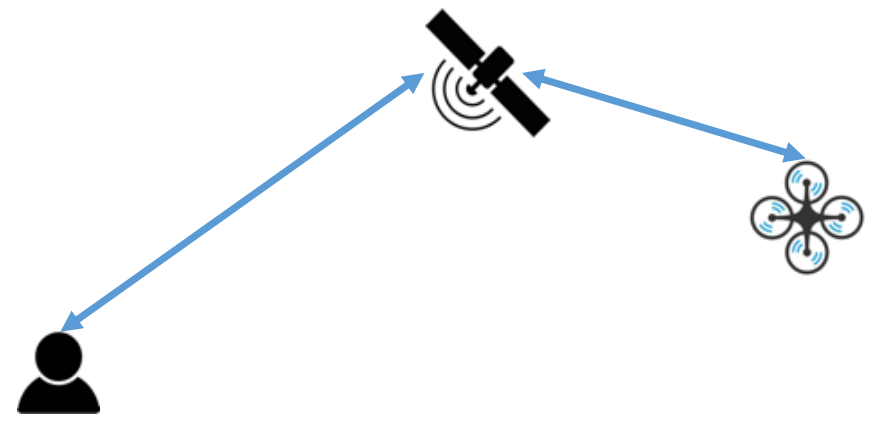

Fig. 3. An illustration of BLOS provided through the Satellite

BLOS communication overcomes LOS data links limitations, including the low coverage range, high power attenuation, and local weather influences. Another reason to move towards SATCOM data links is that LOS data link bands are congested, and rising air traffic growth requires dedicating new radio spectrum for aerial data links. It is important to mention that BLOS includes LOS too, so SATCOM provides the LOS services as well.

Satellite links usually use K band and Ku band. The current trend is towards employing the Ku band for high-throughput satellites (HTS). Ku band SATCOM systems use 11.7-12.7 GHz downlink and 14-14.5 GHz uplink. In the near future, these satellites may provide more than 100 Gbps throughput [53].

### B. *Disadvantages*

Despite all the great benefits of the satellite communication, this technology is an expensive data link, and it becomes cost effective only for high-altitude or at most for medium-altitude UAVs. Hence, SATCOM has not been used for small UASs so far.

One of the main challenges with all satellite communications is latency, due to the far distance that the data packet has to travel. Latency can be defined in two ways: one-way or round-trip latency (RTL). One way is the time that a data packet takes to travel from the sender to the receiver. RTL is the time required for the packet to get to the receiver and a response goes back to its sender.

Due to the high latency of SATCOM data links, real-time remote piloting becomes less practical. In this case, the complete flight plan can be programmed in a chip and the UAV is guided by an autopilot. Meanwhile, a remote pilot may still surveil the aircraft (which is not real-time monitoring) through a control link with about 10 kbps data rate [54]. It is important to note that distance is not the only factor affecting the latency of the SATCOM services. Bandwidth, the load on the network, and the constellation's capacity are some examples of other factors that affect the latency of their services.

Another disadvantage of SATCOM is the high level of propagation loss. Signal attenuation caused by several environmental features (e.g., free space losses, atmospheric losses, signal absorption, and dish misalignment) gets worse as the distance between the transmitter and receiver increases. This requires strong high-power amplifiers to be deployed at the satellites.

SATCOM often suffers from gaps in communication. A constellation of satellites may not cover the whole area of the Earth's surface. At high geographical longitudes (including poles), most satellite constellations are not visible. This is because the motion of the Earth makes launch a satellite into a polar orbit more difficult than launch it into an equatorial orbit. Further, sometimes the satellites are not in view of their ground stations, and the useful bandwidth cuts to about a two-third.

In Table II, the advantages and disadvantages of SATCOM are summarized. Also, its suitability or unsuitability for several UAV applications have been mentioned. However, depending on the available resources (e.g., cost, computational capability of the GS, LOS or BLOS situations, etc.), the constraints might change.

### C. SATCOM Standardization

The Joint Aviation Authorities (JAA) and the European Organization for the Safety of Air Navigation (Eurocontrol) groups are working on the regulation of civilian UAVs, regarding their safety, security, airworthiness and licensing [55]. They have organized a UAV task force whose report contains five main requirements on SATCOM for UAV system type certification. The requirements are as follows:

- National authorities must approve all frequencies being used in UAV operations.
- The remote pilot should supervise the communication link constantly.
- Any failure in the communication links should not affect UAV operations.
- Data links should be protected from electromagnetic interference (EMI).
- The occurrence of communication interruption, random failures, alternative data links, and total path loss of the links should be analyzed in the airworthiness certification process.

TABLE II
SATCOM DATA LINKS FEATURES

| | |
|---|---|
| Advantages | • Enable BLOS<br>• High reliability<br>• Maximum coverage range<br>• Allow high aircraft mobility<br>• Less congested frequency bands |
| Disadvantages | • A higher level of propagation losses<br>• High latency<br>• Expensive<br>• Low data rates<br>• May not be always available |
| Suitable for UAV Applications such as | • Surveillance<br>• Weather services<br>• Agriculture<br>• Remote area coverage<br>• Natural disaster |
| Unsuitable for UAV Applications such as | • Real-time monitoring<br>• Rescue missions<br>• Internet coverage<br>• Construction<br>• Law enforcement<br>• Delivery |

An important challenge in using SATCOM communication for UASs is that there is no standardized frequency band allocated to them in the protected aviation spectrum. In the World Radio Communication Conference 2012 (WRC-12), new spectrum allocations in C band and L band were proposed for Aeronautical Mobile (Route) Service (AM(R)S) for LOS data link used in the UASs. However, the proposal did not consider the BLOS spectrum requirements. The allocated part of the C band, in the range of 5030-5091 MHz, is insufficient to supply both the LOS requirements and the minimum required 56 MHz bandwidth for satellite BLOS UAS data links. Also, there is no satellite currently operating or planned to operate in the C band for use in BLOS UAS data links. Hence, utilizing SATCOM for UAS is still infeasible in the protected C band aviation spectrum.

In the 2015 WRC, the Fixed Satellite Service (FSS) band was considered to provide BLOS data links for UASs. However, the current dedicated bandwidth for the FSS, which is a part of the Ku/Ka-band spectrum, is not in the range of ICAO's protected aviation spectrum [56]. Moreover, some terrestrial fixed service systems operate in the same frequency band that might cause interference. There have been several studies on sharing the spectrum between the UASs and those terrestrial fixed service systems.

### D. Available SATCOM Services

An increasing number of companies are providing satellite communication and are trying to test their services for the UAVs. Unlike the land-based or terrestrial communication, which has been provided by just a few companies offering relatively the same services, there are many different types of SATCOM services. Picking the best service among various available options depends on the application's constraints and requirements of the data link.

#### 1) InmarSAT

InmarSAT was the only SATCOM provider for a long time. After the digital revolution, many other satellite companies providing various types of services appeared.

For unmanned applications, InmarSAT offers a machine to machine (M2M) communication service in L band. This service is a member of InmarSAT's Broadband Global Area Network (BGAN) M2M family provided by three GEO satellites started in January 2012 [57]. The BGAN service provides a throughput of about 492 kbps per user.

InmarSAT also provides critical safety services for UAVs and their associated applications. Some possible UAV applications using InmarSAT service are data reporting for pipelines, environmental and wildlife monitoring, and electricity consumption data.

InmarSAT also offers a hybrid service called Global Xpress

(GX) in combination with BGAN. This service is useful for applications that require no interruption, high availability, and seamless connectivity. The GX satellites offer Ka-band services (in the range of 20-30 GHz) for high throughput and BGAN through its L band service provides high availability. This high level of performance and flexibility of the GX satellites make the total throughput of each satellite to be around 12 Gbps. The GX can supply downlink speeds up to 50 Mbps, up to 5 Mbps over the uplink per user, and both the downlink and uplink of BGAN offer data rates up to 492 kbps per user [58].

The typical latency for streaming service in BGAN system is about 1-1.6s round trip. Hence, the one-way latency is about 800 ms at most. However, only 72 of 89 GX satellite spot beams are available at any time (81%) as they travel over the ocean. So many customers would be periodically and unexpectedly limited to older, very high latency FleetBroadband service. FleetBroadband is a maritime global satellite internet and telephony system built by InmarSAT. The total latency of the FleetBroadband network is in the range of 900 ms to 1150 ms, and the average latency of a GEO satellite is about 500 ms [59, 60]. Therefore, by weighted average, it is expected that the total average latency of a GX system would be around 600 ms. However there is no official document reporting the user experienced GX system latency.

Recently, InmarSAT introduced their new service called InmarSAT SwiftBroadband UAV (SB-UAV) satellite communications service in coordination with Cobham SATCOM. This service can be implemented on low-altitude UAV to provide a satellite communication link for BLOS applications. However, this service suffers from large latency.

**Conclusion:** InmarSAT GX offers services suitable for UAV missions in which seamless communication is essential through their hybrid GX/BGAN service. Applications such as surveillance and delivery are two examples. The BGAN service operating in L band is a proper data link in the matter of supporting mobility due to the Doppler frequency challenge. On the contrary, latency is a disadvantage of InmarSAT that makes it unsuitable for applications that require low latency communication such as real-time monitoring.

*2) Iridium NEXT*

Iridium's first goal was to build a space-based type of cellular network stations through 66 satellites. As a common challenge for all SATCOM providers, Iridium has a lapse in coverage of about 4% of the time.

Iridium NEXT is the second-generation of Iridium satellites for telecommunications providing worldwide narrowband voice and data services. The constellation offers services in L band for mobile users, supplying data rates up to 128 kbps per user [62]. Iridium NEXT has recently (July 2018) extended its current services by lunching another 10 satellites.

In the Iridium NEXT, the throughput increased compared to the first-generation constellation. However there is no official document reporting the throughput [61, 62]. This set of satellites supplies the users with a fast, secure and comparatively lower latency communication links [58, 63]. According to [64], packet delays in the first Iridium generation had the average of 178 ms. With the enhanced performance in Iridium NEXT, the delay is below 40 ms [65].

Iridium services have been used widely to provide satellite communications for UAV hosted personal communication services (PCS) for warfighters using low cost, battery powered handsets. Each UAV acts as a relay station to extend the coverage. UAVs provide BLOS data services to about 1000 handsets in its coverage area [66]. These data links are suitable for low and medium endurance UAVs [4].

**Conclusion:** Iridium's satellite services operating in the L band would help with Doppler shift challenge in UASs, due to its lower frequency compared to Ka band. This service has been widely used for military handsets communications and shown a good performance in large coverage areas. The benefit of LEO satellites in this constellation is its relatively lower latency communications. However, this comes with the higher cost and bigger antennas which make it unsuitable for small UAV applications.

*3) Globalstar*

Globalstar is an LEO satellite constellation that operates in both S and L bands. The second generation of the Globalstar constellation has 24 LEO satellites. Its launch started in 2010 and was finished by early in 2013 [67]. Globalstar is known as an Iridium-like service; it has a delay of about 40 ms [65]. There is no official document reporting the throughput per satellite. The average data rate provided by the system is approximately 7.2 kbps per user [68].

Globalstar and ADS-B (Automatic Dependent Surveillance-Broadcast) Technologies have been cooperating for several years for aviation communication services. They provide a simple and low-cost satellite-based ADS-B system called ADS-B Link Augmentation System (ALAS). The main goal is that when the aircraft is not in the LOS of the ground station, the Globalstar satellites provide an NLOS communication link for the ADS-B signals. This system guarantees a highly reliable NLOS air traffic management (ATM) system. In other words, this service extends the ADS-B coverage into BLOS areas with almost no performance degradation in non-satellite-based communications [69]. Also, it does not add any interference to the other aircraft's normal transmissions. ADS-B is discussed further in Section VI.

Recently, these two companies, Globalstar and ADS-B technologies, in coordination with NASA Langley Research Center, integrated the ALAS service for UAV applications. A Cirrus SR22 aircraft was used as a test vehicle and flown remotely from the ground. The test results indicated that the system delivers a constant rate communication link between the UAV and the satellite with only a few breaks and quick reconnections [70].

**Conclusion:** Similar to Iridium services, this SATCOM communication system also provides robust data links against Doppler shift operating in S and L bands. With high mobility and relatively lower latency services, Globalstar is a potential data link for a wide range of UAV applications.

*4) Orbcomm Generation 2*

Orbcomm Generation 2 (OG2) is the second generation on Orbcomm constellation. The constellation uses very high frequency (VHF) band and frequency hopping to avoid interference in this crowded band. The satellite average latency has been reported as under 1 minute in almost all on-ground operations [71]. OG2 is dedicated to M2M communications. This constellation of satellites consists of 18 satellites, with the total 57 kbps throughput per satellite [72]. The data rate per user has not been reported.

Orbcomm's services are mostly designed to work in unmanned environments for remote tracking and monitoring of oil and gas extraction and distribution [73]. Orbcomm also provides low power Internet of Things (IoT) services and M2M communications that can be used in multi-UAV networks. It has established a combined robust network consisting of satellite service and terrestrial cellular network along with dual-mode network access. This helps provide a flexible communication system to accommodate the user's demands.

**Conclusion:** OG2 provides links with low power communication, which is desirable especially considering the SWaP limitations of UAVs. Their hybrid service in combination with the cellular network can satisfy a wide range of service requirements based on UAV's specific task; however, it will not be a pure SATCOM service. Their low-frequency operational band makes these services suitable for a wide range of UAV application. Their operating VHF band supports a high level of mobility without facing Doppler shift challenge. Even they have a relatively smaller distance through their LEO satellites, 1-minute latency is not proper for real-time applications.

*5) OneWeb*

The main motto of founding OneWeb was to provide affordable internet services in the current under-developed regions [74]. Satellite network provided by OneWeb, formerly known as WorldVu, will consist of 648 LEO satellites to provide a broadband global internet service by the end of 2019. The satellites would operate in Ku band, 12-18 GHz range of the radio spectrum. The throughput of each satellite is anticipated to be about 6 Gbps.

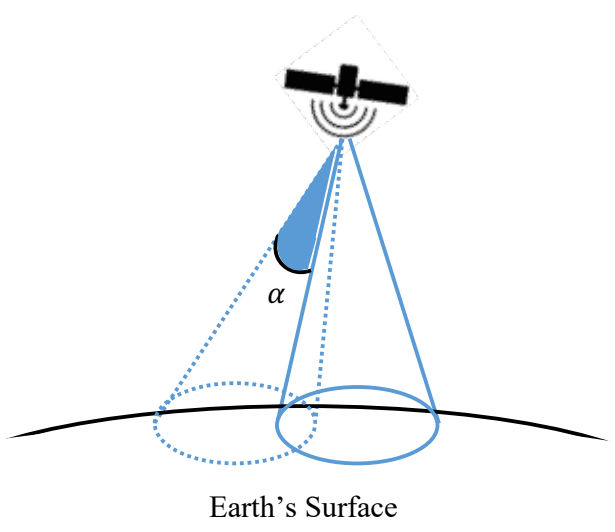

Fig. 5. OneWeb Progressive Pitch

OneWeb suggested a new technique called "progressive pitch" to be implemented in their constellation. In this method, the satellites will be slightly turned occasionally to avoid interference with other Ku band satellites in GEO. This is shown in Fig. 5, where the lobe has been moved to the left by α angle.

OneWeb will support variable data rates, depending on the instantaneous modulation and coding scheme. It is expected to be at least 10 Mbps data rate per user, however; the exact data speed is not available yet [75, 76].

OneWeb's constellation benefits from the main latency advantage of LEO satellites, which have a low RTL compared to higher orbits. OneWeb services could have latency as low as 50 ms, while the latency of a typical office LAN or ADSL (asymmetric digital subscriber line) connection is in the range of 15-100 ms.

OneWeb will support UAV operations over the Arctic, an area recently opened to maritime lanes, but it is beyond the reach of GEO satellites. The Arctic is the polar region located in the northernmost part of Earth.

**Conclusion:** OneWeb services would suffer less from interference due to their progressive pitch technique, which is a bonus for SATCOM data links used in UASs. Further, along with their relatively lower latency communication links, this SATCOM service would be suitable for most critical satellite-based UAV missions, such as network coverage for remote areas. However, since their operating band is in the Ku band, they are not able to support applications that require high mobility due to the Doppler shift problem. Also, it is expected that this SATCOM service comes with high costs due to the number of satellites and maintenance complexity.

*6) O3b Networks*

O3b Networks offer SATCOM services deploying high speed and medium latency satellites that deliver internet services to remote areas such as Africa, South America, and Asia. The company was founded in 2007 [77].

O3b Networks introduced their latest product, "O3b satellites," constellation containing 12 satellites, while they are planning to extend to 20 satellites by 2021. O3b service has a round-trip latency of approximately 132.5 ms for data services [78]. High-performance satellite terminals support service rates up to 24 Mbps [79]. This constellation operates in the medium earth orbit (MEO). The MEO satellites operate in altitudes between 2,000 km and 35,000 km. The total throughput is 16 Gbps per satellite [80].

For UAV applications, O3b network can be used as an IP-based optimized satellite system solution [81]. SES Government Solutions, a company that now owns the O3b network, offers robust communication capabilities for remotely piloted aircraft using O3b satellites. It also provides flexible operations for advanced remote-controlled sensor platforms [82].

**Conclusion:** O3b can help UAV applications by providing network coverage in the remote areas. Even though the service is very reliable and robust, it suffers from a high level of latency. Their operating frequency band does not support a high level of mobility either. As a result, they are suitable only for low mobility UAS applications with no latency constraints, where a high level of reliability is required, such as secure data collection.

TABLE III
ACTIVE COMPANIES PROVIDING BLOS SATCOM FOR UAS APPLICATIONS

| **Company** | InmarSAT | InmarSAT | Iridium | Globalstar | Orbcomm | OneWeb | O3b Networks | SpaceX |
|---|---|---|---|---|---|---|---|---|
| **Products** | BGAN | GX | Iridium NEXT | Globalstar | OG2 | OneWeb | O3b | SpaceX Satellites |
| **Operational Orbit** | GEO | GEO | LEO | LEO | LEO | LEO | MEO | LEO |
| **Number of Satellites** | 3 | 3 | 66 | 24 | 29 | 648 | 12 | 4000 |
| **Bandwidth** | L band | Ka band | L band | S & L band | VHF band | Ku band | Ka band | Ku band |
| **Data Latency** | 800 ms | 600 ms | 40 ms | 40 ms | 1 minute | 50 ms | 132.5 ms | 150 ms |
| **Throughput Per Satellite** | 800 Mbps | 12 Gbps | N/A | N/A | 57 kbps | 6 Gbps | 16 Gbps | 50 Gbps |
| **Total Throughput** | 2.4 Gbps | 36 Gbps | N/A | N/A | 1.7 Mbps | 4.2 Tbps | 192 Gbps | 200 Tbps |
| **Data Rate Per User** | 492 kbps | 5 Mbps | 128 kbps | 7.2 kbps | N/A | N/A | 24 Mbps | N/A |
| **Date of Service Available/Expected** | Jan. 2012 | Dec. 2015 | Jan. 2017 | Feb. 2013 | July 2014 | 2019 | Nov. 2014 | 2020 |

*7) SpaceX*

SpaceX has declared to build internet services from the space by implementing a network of 4,000 small and low-cost LEO satellites in the Ku band spectrum, promised to be fully functioning in 2020. SpaceX is cooperating with Google to construct an LEO satellite constellation, which will provide low-latency and high-capacity internet services worldwide [83]. The total promised throughput is up to 200 Tbps.

SpaceX plans to improve the latency by placing the satellites in a lower earth orbit at 650 km and also having space connections among the satellites [84]. By this strategy, the latency would decrease from 150 ms to 20 ms, which is about the average latency of a fiber optic cable internet for home services in the United States [85, 86].

**Conclusion:** Even though SpaceX has not been employed in civilian UAVs, it has a great potential. Due to the promised low latency service (if it is successfully implemented), it can be tested out for near real-time monitoring. UAV applications such as weather services, agriculture, and delivery operations can benefit from a SpaceX data link communication. However, claiming to cover all the internet users with the promised throughput might not seem very practical considering the future drastic number of users. Some UAV missions demand very high throughput, especially when video streaming is needed.

*E. Related Research*

Finding a proper solution to the problem of sharing the same spectrum with FSS in Ku/Ka band have been studied in [87, 88]. In these works, the primary focus is to achieve an efficient BLOS satellite data link for the UAVs. However, as concluded, a specific regulation for SATCOM is still needed.

From another perspectives, there are other on-going research works in this area. For instance, designing a satellite-based antenna system operating in the Ka band between the UAV and the remote pilot have been studied in [89, 90]. The proposed onboard satellite antennas designed for UAVs are low-profile and broadband antennas that are very small in dimensions and operate in a wide range of frequency spectrum. These two main features are very useful and essential for designing small UAVs.

Improving the UAV situational awareness has been studied in [91]. The proposed solution is based on establishing a collaborative mechanism between UAVs using satellite communication. In the paper, UAVs are called unmanned satellite vehicles (USV). The positive aspects of using a swarm of collaborative USVs in a small area are analyzed. For instance, the USVs are able to finish their mission autonomously without any human interactions. Several situational awareness missions such as resource searching mission, fire detection mission, critical infrastructure surveillance and warning detection are considered.

For collision avoidance, utilizing satellite-based radar for UASs has been studied in [92]. A modified Automatic Dependent Surveillance-Broadcast (ADS-B) system is used. The main objective of the research is to ensure the safety of NAS through the co-existence of UAVs with other aircraft. The proposed ADS-B satellite radar satisfies this objective by sharing the situational awareness information among all the aircraft. This area of research is essential due to the requirements of UAS integration into NAS and for faster improvements in the future ADS-B systems.

Skinnemoen [93] has studied the challenges of using SATCOM for UAVs. In other words, the main focus of this research work is on live photo and video sharing using satellite communication. They have used InmarSAT BGAN service in their work.

Some primary studies on using small satellite antennas in future have been tested recently on UAVs in [94]. Simulated SATCOM service providers are used to mimic the behavior and feature of a real satellite data link. However, in the built testbed, the simulated small SATCOM antenna is able to communicate only in a half-duplex mode, so it might not be a proper representation of a real-world case. Hence, more research work is required in this area, even though these primary studies are a huge step.

## *F. Summary*

Even though satellite communication is expensive to use in civilian UAVs, it provides a large coverage area for high-altitude unmanned aircraft. As discussed in this section, there are a wide-range of services operating in various frequency bands with different ranges of data rates, and link latencies that can be chosen based on the application constraints. Even though the current SATCOM services do not offer high throughput services per aircraft, emerging the HTS satellites may be promising in future [95].

If SATCOM BLOS communication is used in the UAS, it is better to use a service from LEO constellations due to its lower communication delay. However, in the case of fully autonomous missions, the whole flight plan can be programmed in an on-board processor and so the delay may not be an issue. However, this method is not proper for mission-critical tasks, since in case of any unexpected change in the plan, the pre-programmed UAV will not be ready.

Almost all the satellite-based service providers have started testing their data links for UAV applications. Inmarsat, Iridium, and Globalstar have built satellite-based products specifically for aviation, as it was mentioned earlier.

The general data link information regarding the discussed SATCOM services is summarized in Table III. Their main features in terms of different constraints is summarized in Table IV. Since SpaceX is not operational yet, it is not included in this table. It should be noted that if a SATCOM service is not marked for a specific feature, we are not excluding that service.

TABLE IV
MAIN FEATURES OF EACH SERVICE BASED ON UAV'S COMMUNICATION REQUIREMENTS

| SATCOM Services | Relatively Lower Latency | More Robustness & Reliability | Higher Level of Mobility | Better Regarding SWaP | Higher Availability |
|---|---|---|---|---|---|
| BGAN | | | X | | |
| GX | | | | | X |
| Iridium NEXT | X | | X | | |
| Globalstar | X | | X | | |
| OG2 | | | X | X | |
| OneWeb | X | X | | | |
| O3b | | X | | | |

# V. CELLULAR AVIATION

In recent years, telecommunication providers have shown a great interest in adapting their technologies to provide services for cellular aviation. Qualcomm, an American cellular equipment company, is leading a trade group focusing on the technologies dedicated to unmanned systems and robotics industry. They predict that UAVs will benefit the United States' economy by $13.6 billion in the first three years after they are standardized to operate in the NAS [96].

In the following subsections, advantages, disadvantages, related research and application of the 4th and 5th generations of cellular technology generations are highlighted.

## *A. Advantages*

To catch up with the tremendous growth in the wireless space related to unmanned applications, it is helpful to take advantage of the existing allocated bandwidth to the mobile wireless communications. Cellular spectrum capacity can be scaled up with some additional planned spectrum to be used in the UAS aviation sector. It has the potential of providing a higher level of reliability, robust security, and seamless coverage in the UAS operations. All these improvements along with enabling the BLOS data link by cellular service serve the UAS applications and requirements.

Exploring cellular data links sounds promising to solve satellite limitations. UAS BLOS operations are limited due to the shortage of suitable satellite bandwidth. Further, satellite communication links alone may not satisfy the increasing demand of small UAS operations in future.

Cost and delay are other challenges that can be overcome by cellular technology. In the cellular communication, link redundancy may be possible, which means if one link fails or operates poorly, the system can switch to a better link. This provides UAS with resilient and seamless connectivity which is desirable for several applications.

Cellular connectivity can supply a safe and autonomous flight by providing dynamic optimal flight plans, based on the current UAV location and situation. The low latency provides the remote pilots with almost real-time services for tracking and determining the safest routes for the UAV. Further, cellular data links can support real-time video streaming between the aircraft and the pilot.

## *B. Disadvantages*

Even though cellular communication for UAS applications sounds promising, it comes with several disadvantages. Cellular networks have been used for various applications, and with the increase of mobile applications use, the allocated bands get congested. This situation gets worse in crowded areas. Therefore, there is a need for specialized bands to be allocated for UAS applications to meet the requirements. On the contrary, as was discussed before, SATCOM bandwidth availability is higher and less congested.

Another disadvantage of cellular networks compared to SATCOM services is that SATCOM services offer longer range coverage. Cellular network towers have short range coverage area and need several handovers during UAV missions. Plus, they are not available in some rural or remote areas. Thus, in some cases, providing the BLOS communication will be limited to the SATCOM services due to their larger coverage.

Different weather conditions may affect the quality of the cellular service as well. The proper solution for this problem is usually through increasing the transmission power or sending redundant copies of the data, which was already explained as diversity techniques. Increasing the transmitted power is usually considered as wasting power and causes interference with others. Further, it constrains the SWaP limitations.

One other critical challenge is that the cellular infrastructure is not designed for aviation communications. Most of the antennas transmit signals towards the ground and not upwards.

This can cause loss of connection even when the UAV is flying at high altitudes [97]. To support UAVs, the signal transmission patterns will have to be changed, so that some of the side lobes point upwards. However, this change will not happen until there is significant UAV traffic to justify the investment. Regardless of the primary purpose of cellular networks, it can still be used as a data link for UAVs shared with the other cellular users.

There are different aspects that must be considered based on the end nodes of communication while using cellular aviation for the UAS. These include resource allocation, latency requirements, bit rate, etc. For instance, to design data links for communications among UAVs, resource allocation (e.g., bandwidth, fairness requirements and transmission time) should be considered. For the data link between the remote pilot and the UAV, constraints such as latency, bit rate, and the loss rate are main factors depending upon the degree of autonomy. High bandwidths might be required if these constraints are tight.

### C. *Cellular Aviation, 4G, and 5G*

As the mobile communication evolved to the fourth generation (4G), data rate, latency, throughput, and interference management improved significantly. Also, 4G or LTE enabled data links are dynamic and can be configured based on UAV requirements. Thus, they are considered a potential candidate for several UAS applications.

As the fifth generation (5G) is approaching, many promises have been made to improve the 4G services by delivering ultra-high reliability, ultra-high availability, incredibly low-latency, and strong end-to-end security.

Some of the significant features provided by 5G have been summarized in Fig. 6. 5G has promised to increase efficiency by enabling all these features with a lower cost for a wider area. It is planned to use 5G services to support public safety using bands above 24 GHz and employ UAVs or robot-based surveillance systems to provide remote monitoring [98].

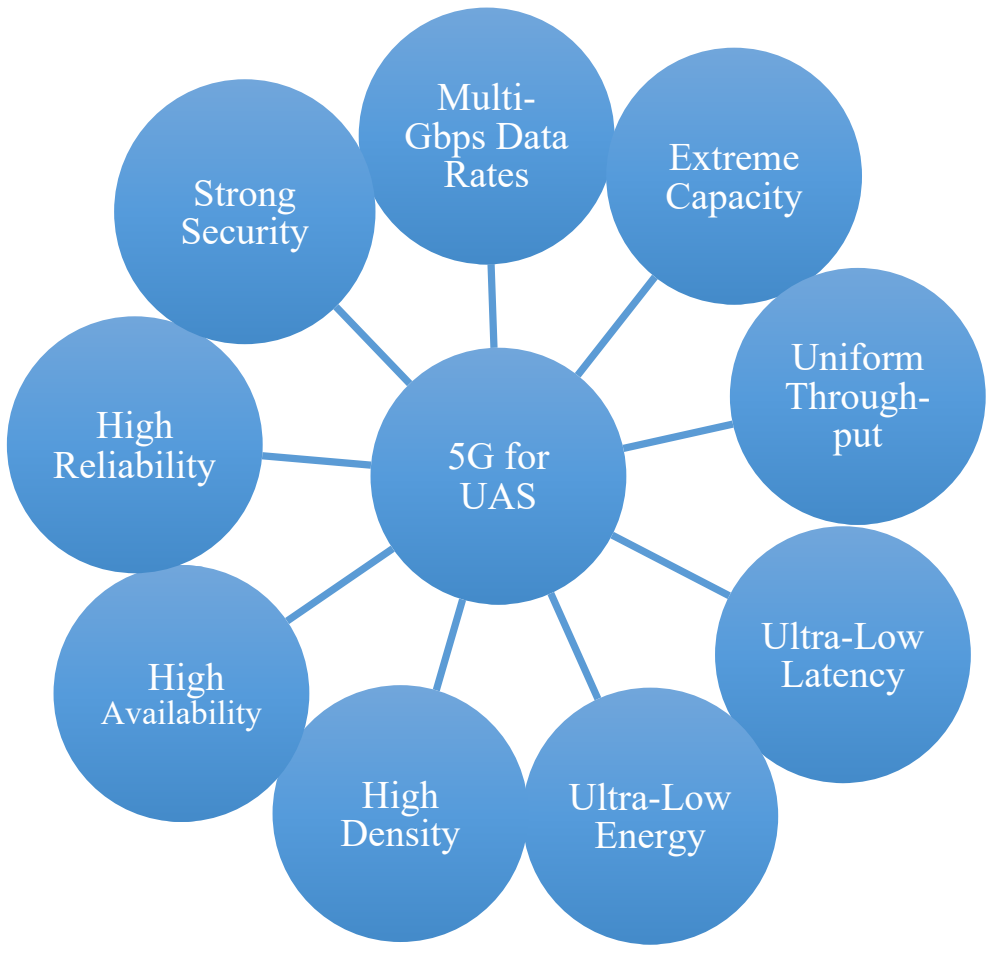

Fig. 6. An illustration of the features enabled by 5G

In the following, we will discuss important features of 4G and 5G services that make them suitable for cellular-based UAV applications.

#### 1) *Availability*

The data links used in UASs requires high availability so that the remote pilot can have constant access to the UAV. Coverage range of cellular towers is not very high. Hence, in the missions that UAVs need to travel a long distance, each UAV will be served by multiple cell towers. Properly optimized handover mechanisms need to be planned to increase the coverage range with no lapse in the communication.

One of the exceptional characteristics of 4G LTE is the Coordinated Multi-Point (CoMP) base station technology. In this technique, two or more base stations coordinate transmissions and receptions to the user to improve the availability, especially at the cell edge. Having a high-quality data link to the base station even at the cell edges can improve the availability and performance and avoid any collision caused by the poor communication. CoMP is important to achieve the required SINR (Signal to Interference plus Noise Ratio) of the system. Further, CoMP will enable QoS features and enhance the spectrum efficiency [99].

On the other hand, in cellular communication, the coverage and availability are also dependent on the node density of that particular area. Hence, cellular networks might not be an efficient choice in highly dense or technologically advanced areas [100].

Another important factor regarding the cellular networks is the reuse distance. Reuse distance means that the cellular network can reuse the frequencies in specific distances based on the inference level. This feature increases both the availability and capacity of the network. The reuse distance is dependent on the tower’s cell radius and the number of cells per cluster in a specific area. However, with increasing capacity, the reuse distance becomes very short, the reuse cells start to overlap with each other, causing interference, thereby SINR decreases significantly.

#### 2) *Throughput and Data Rate*

The throughput provided by the 4G network has improved 10 times more than 3G technology, which is relatively sufficient for video services. However, it is promised that 5G would offer a much higher level of throughput that would be uniform with no lapse in connection. This will improve the UAV’s video-based applications even further.

The Federal Communications Commission (FCC) has planned 5G mobile networks to be implemented in specific frequency bands in the tens of GHz, called "millimeter-wave" bands. Millimeter wave would enable 5G to have gigabit-per-second data rates, which supplies the UAVs with previously mentioned ultra-high-resolution video communications. However, because the frequency is high, signal propagation becomes a challenge and should be taken care of by the carrier providers. Also, these frequencies do not penetrate into buildings as easily as the lower bands [101].

Exploiting 5G as the data link for UAS and the increased throughput will also enhance the direct UAV-to-UAV communication in multi-UAV networks. UAVs with high data rate data links employed in a mesh network acting as flying relays to help the data exchange between terrestrial users has been studied in [102-105].

### 3) Latency

Through ultra-low latency 5G networks, new mission-critical services will be possible in the UAV application domain. That means 5G communication capabilities may be expanded beyond human constraints, in latency and reliability. In UAV-based mission-critical applications, it is crucial to have seamless connectivity, and failure is not an option [96].

The latency of 4G networks has improved to about half of the 3G technology and is about 50 ms. The expected latency of the 5G is promised to be less than 1 ms. This level of ultra-low latency enables designing proper data links for unmanned and automated technologies while guaranteeing the mission's safety.

## D. UAVs for Cellular Services

UAVs can also be very useful for cellular services (e.g., in case of emergency or disaster) to perform as a flying base stations (BS). In cellular communications, a wireless connection to the public telephone network is established through the local cell tower. During a disaster, these towers may lose their functionality. This leads to the loss of communication in the affected regions which could lead to further disasters. In such situations, constant coverage and communication are vital for public safety. In this case, UAVs can establish instant connectivity by implementing cognitive radio in a multi-UAV network to form a wireless mesh network with devices in the affected area.

Cognitive radio concept is based on changing the spectrum access dynamically for the opportunistic utilization of licensed and unlicensed frequency bands in a specific area. Since cognitive radio networks are infrastructure less and spontaneous, they are very suitable in disaster situations. There are detailed investigations in papers [106-108] on employing cognitive radio technique in UAS networks. A comprehensive survey of cognitive radio for aeronautical communications is provided in [109]. They also discuss the significant performance improvement that cognitive radio brings for UASs. Cognitive radio would also help the problem of increasing number of civilian UAVs facing spectrum scarcity.

UAVs can also form an ad-hoc network to replace the malfunctioning tower in the cellular network. Once set up, the UAVs can act as mobile base stations and start routing traffic to and from the cell tower. However, due to their limited power sources, using UAVs is a temporary solution while trying to restart communication through the permanent networks [110].

In addition to that, similar techniques can be used in remote or rural areas that lack cellular towers. A similar UAV-based BS concept can be applied to provide temporary cellular connection and internet access for the users to cover the area.

## E. Related Research

This section is divided into two parts. First, we discuss the research works considering the UAVs as users of cellular networks. Next, we highlight the research case studies focusing on deploying UAVs as flying BS to provide assistance for the cellular networks.

### 1) UAVs as Users

The applicability of using 3G and 4G mobile communications for UAVs' data link has been studied in [111]. The results of the paper show that the Long-Term Evolution (LTE) and UMTS network provide secure, low latency, and high throughput data exchange. These features are critical in the UAV applications. On the other hand, level of readiness and issues of 5G cellular network for drones are looked into in [112].

In other research, LTE-based control and non-payload communication (CNPC) network for UAVs is investigated [113]. Security aspects are also studied since security is highly important for command and control data links. Any failure or malicious attack can jeopardize the whole mission.

UAS civil applications using cellular communication network has been studied in [114]. Different technologies such as EDGE, UMTS, HSPA+ (High-Speed Packet Access Plus), LTE, and LTE-A (LTE Advanced) have been investigated. Some experiments on radio propagation are presented as well. Wide radio coverage, high throughputs, reduced latencies, and large availability of radio modems are mentioned as advantages of using cellular communication for UASs.

Some discussion on an integrated UAS CNPC network architecture with LTE cellular data link are presented in [115]. A new authentication mechanism, key agreement protocol, and handover key management protocol are also proposed. Providing authentication security policy is very useful in sensitive UAV applications such as delivery, industrial inspection, monitoring, and surveillance. Since it is important to make sure that only authorized users are able to access the data.

The potential and challenges of integrating the UAVs to cellular networks as aerial users are studied in [116]. Some primary studies regarding different UAV heights have also been shown in the paper.

Using cellular data links for UAVs at the same time with serving the ground users is the main focus of [117]. Comprehensive analysis of current and next-generation cellular networks are studied. The current traditional topology is based on single user mode, which means one user is served per frequency-time resource at each time, whereas in next generation, multi-user massive MIMO (multiple input multiple output) BSs are available. The latter means at each time, multiple users will be served per frequency-time resource.

### 2) UAVs as BSs

Investigating the performance of a UAV acting as a mobile base station is provided in [118]. The main objective of the paper is to improve the coverage and connectivity in a specific area in which users are cooperating using device to device (D2D) protocol. There is a comprehensive study on medium access control (MAC) design for UAV-based ad hoc networks using full-duplex and multi-packet reception (MPR) abled antennas [119]. In their scheme, each UAV uses code division multiple access (CDMA) technique to model the MPR. The simulation results show that the idea of combining these two capabilities (full-duplex radios along with MPR) will significantly enhance the performance of UAV-based ad hoc networks.

In a comprehensive tutorial paper, [120], the potentials and advantages of using UAVs as aerial base stations in cellular networks are studied. The authors discuss the challneges, future research directions, and statistical methods for analyzing and improving their performance. They also have presented their study results, [40], on energy consumption efficiency regarding optimal localization and number of deployed drones.

As mentione before, the cellular networks were not designed originally for flying users. In [121], the focus is on trying to tackle this challenge by desinging a UAV-based cellular network specifically designed to serve the drone applications. Through optimizing the three dimensional placement and frequency planning, the proposed system showed a significant improvement in the communication delay for the flying drone users.

In [122], a dense urban scenario is considered where the existing cellular network falls short in satisfying the users' requested QoSs. A UAV assistant BS is designed in the network to handle the users that could not be served by the main network due to overload. A joint versatile configurations is proposed based on locating the drone in a proper three dimension and the optimal incentive offered to each user. It might be useful to mention that there exist an older study, [123], that has initiated this idea of using UAVs to assist the existing terrestrial BSs through optimizng the number of deployed UAVs and their location.

### F. Summary

Even though cellular infrastructure is not primarily designed for aviation, it is considered as a potential and cost-efficient choice, especially for BLOS missions. High cost and lack of bandwidth for satellite-based unmanned aviation are also the reasons to move towards using cellular data link communications for UAV mission.

Cellular aviation has some significant benefits over other techniques, such as high reliability and data rate. Using light weight (100-150g) cellular device on small size UAVs are practical and cheap solution from the hardware point of view.

Using UAV-based mesh networks in case of disaster where the cellular towers are damaged or in rural areas where cell towers are not available to provide connectivity among the users in the affected area is an interesting area for researchers.

Cellular communication in aircrafts has been employed since the second generation of mobile communication. Providing direct internet access to the aircraft was an important milestone in aviation. However, the use of cellular communication is only possible for UAVs flying under 125 m above ground level due to decreasing signal to interference and noise ratio (SINR). At higher altitudes, satellite data links are mostly used.

4G LTE makes it possible to take advantage of cellular networks in regular UAV applications. However, mission-critical applications need still higher levels of reliability, availability, low-latency, throughput, and security. These features are promised to be available through the next generation of mobile technology, 5G. A summary table on several characteristics of all cellular generations has been provided in Table V.

TABLE V
DIFFERENT GENERATIONS OF CELLULAR COMMUNICATION TECHNOLOGY

| Generation | 2G | 3G | 4G | 5G |
|---|---|---|---|---|
| Data Rate Per User | 64-144 kbps | 144kbps-2 Mbps | 2 Mbps-100 Mbps | up to 1 Gbps |
| Max System Bandwidth | 25 MHz | 25 MHz | 100 MHz | 500 MHz-1 GHz (See Note) |
| Switching Type | Circuit switching | Packet & Circuit switching | Packet switching | Packet switching |
| Latency | 692 ms | 212 ms | 50 ms | 1 ms |
| Band Type | Narrow-band | Wideband | Wideband | Not defined yet |

Note: LTE allows up to 5 carriers of up to 20 MHz to be aggregated. 5G allows up to 16 carriers, each up to 100 MHz, to be aggregated. 5G allows aggregating the LTE carries as well. While the total theoretical bandwidth is large, any implementation will be constrained by the available practical bandwidth which is currently very limited.

## VI. AVIATION COMMUNICATION STANDARDS

International Civil Aviation Organization (ICAO) has constrained control and communication data links to operate in the protected aviation spectrum and follow the International Telecommunication Union (ITU) restrictions. The reserved services introduced by ITU for flight regulation and safety include Aeronautical Mobile (Route) Service (AM(R)S) and Aeronautical Mobile Satellite (Route) Service (AMS(R)S) [124]. The ICAO tries to make sure that the allocated spectrum for UAS data links and command and control are either in AM(R)S or AMS(R)S, while they do not impact the performance of other aerial systems.

In this section, we discuss some of the aviation standards which are being used for UAVs as well to secure the flights in the NAS. We study their technical details and current related research. These standards include AeroMACS, L-DACS, ADS-B, IEEE 802.11, VDL Mode 2, CPDLC, and SWIM.

### A. AeroMACS

Aeronautical Mobile Airport Communications System (AeroMACS) is based on WiMAX standard, IEEE 802.16. This standard is defined for the physical layer and the MAC layer of the ground-to-aircraft and aircraft-to-aircraft communications at airports [125]. AeroMACS was developed by the Radio Technical Commission for Aeronautics (RTCA) and then proposed at WRC-2007 (World Radiocommunication Conference 2007).

It can provide different QoS based on various network constraints such as error rate, throughput, time delay, and resource management. Also, this standard is flexible regarding scalability for both large and small areas, with cell sizes up to 3 km. This standard operates in C band of the protected AM(R)S spectrum (5091-5150 MHz). It provides data rates up to 54 Mbps per system. Standardization of AeroMACS by RTCA is complete. It is being used in public trials in the United

States [126].

The FAA considers AeroMACS as an important element of the future communication system. Some current applications of AeroMACS include airline operational communications (AOC) messaging, ground traffic control, controller pilot data link communication (CPDLC) messaging, weather forecast information, ATM, airport operations.

AeroMACS bandwidth may need to grow over time, and the allocated RF spectrum to AeroMACS will have to increase to be able to satisfy the unmanned aviation needs [127].

In [128], a large number of flight tests have been done to provide a seamless connection through smooth handovers among three future data links: AeroMACS, VDL Mode 2 and BGAN. The project is called SANDRA, and the main goal is to achieve flexible and scalable network connectivity.

Some AeroMACS services can benefit from a flexible asymmetric ratio of the number of OFDMA symbols assigned to downlink (DL) and uplink (UL) channels. This concept is studied in [129]. Their research is based on the fact that AeroMACS's TDD framework supports different shares of throughput between DL and UL, which can be beneficial for many UAV applications. They provide a comprehensive analysis of different DL/UL symbol ratio. The examined ratios are based on the cell constraints and data rate requirements. The main focus of the paper is on real-time applications such as video surveillance and sensors.

Primary studies on how to apply IEEE 802.16j multi-hop relays to the AeroMACS prototype to enhance its capacity and flexibility are discussed in [130-132]. The proposed method increases the ground station capacity and provides the transmitters with transparent and non-transparent relay modes. This technique reduces the interference.

**Conclusion:** AeroMACS is a good candidate for UAV communications due to its scalability and flexibility. Such a flexible scheme can be adapted based on the UAV mission requirements. Supporting different shares of data rates for the DL and the UL would compensate any resource limitation. However, its limited bandwidth may need to grow to be able to support both manned and unmanned aircraft. Further, its limited coverage area must be extended using several GSs for UAV applications requiring larger coverage. Another minor point is that, since this standard is primarily designed for fixed or stationary users, the mobility support is not that high.

### *B. L-DACS*

Eurocontrol organized a joint European and American project, called aeronautical Future Communication System (FCS), in 2004. This project was initiated to come up with a solution to the growing demand for the aeronautical communications. L-band Digital Aeronautical Communication Systems (L-DACS) was proposed to solve the problem of the continental aviation communication. L-DACS is supposed to be a part of the Future Communication Infrastructure (FCI) program for L band, 960-1164 MHz, under the AM(R)S by the ITU. The two proposed L-DACS technologies are [133]:

- L-DACS1, which is adapted from IEEE 802.16 standard (WiMAX) developed as the public safety communications standard at the Association of Public Safety Communications Officials (APCO) Project 34 (P34). This standard is the next generation of the Broadband Aeronautical Multicarrier Communication (B-AMC) standard, and the Telecommunications Industry Association standard 902 (TIA-902).
- L-DACS2, which is similar to the GSM, is based on the All-purpose Multichannel Aviation Communication System standard (AMACS) and the L band Data Link (LDL) using GMSK (Gaussian Minimum Shift Keying) modulation.

L-DACS1 offers interoperability among services so that they would share the same hardware that provides navigation and surveillance. L-DACS1 performs more efficiently compared to L-DACS2 and is considered as an almost mature technology [134]. Data transmission in L-DACS1 happens in full-duplex which means transmissions happen in both directions (i.e., downlink and uplink) simultaneously. Whereas, in L-DACS2, uplink and downlink transmissions take place alternatively, in a half-duplex method. We refer the readers to [135] for more information on these two techniques and their frame structures.

Several research papers exist on L-DACS1 such as [136-138]. A new multicarrier communication system operating in L band based on filter-bank multicarrier (FBMC) was investigated in these papers to enhance the advantages of L-DACS1.

Of the two versions, only L-DACS1 is now active. It is being considered to be a part of the NextGen as a multi-purpose aviation technology for CNS. L-DACS1 also offers interoperability among ATM services (e.g., navigation and surveillance) [134].

Conclusion: L-DACS1 is an aviation standard, being considered for UAVs. The main advantage of this standard is its operating frequency, which helps the system support high level of mobility.

### *C. ADS-B*

Automatic Dependent Surveillance-Broadcast (ADS-B) is a standard developed by the FAA. Current ADS-B systems will evolve to a next generation, called "ADS-B Next." The current systems work on 978/1090 MHz, but the next generation will be centered initially at 1030 MHz for more robustness and better efficiency. Complete maturity and equipping the ADS-B Next systems are expected to happen in 2025 as a part of the FAA's NextGen program [139]. Upon deploying ADS-B Next, additional bandwidth would be provided through spectrum re-allocation of the 1030 MHz band.

In the ADS-B sense and avoid systems, there are two communication links with different frequencies. Aircraft operating below 6 km use 978 MHz Universal Access Transceiver (UAT), and aircraft operating above that height deploy 1090 MHz Extended Squitter (1090ES) data link. Currently, ADS-B systems mostly operate in a single frequency link, because operating in two different frequencies will cause compatibility issues for communication between different aircraft. Further, trying to address this problem may not be cost-efficient due to the SWaP and budget limitation. However, in manned aircraft, dual frequency ADS-B systems are widely used. The dual frequency scheme can be implemented in a switching manner [140]. ADS-B systems suffer from the data loss for distances above 280 km between the aircraft and the ground base, and this data loss starts increasing almost linearly after 50 km. Hence, the aircraft might not be able to get too far from the nearest base station [141].

UAT is planned to be implemented on all aviation aircraft operating at or below Class A altitudes in the NAS. UAT supplies the aircraft with traffic information, called Traffic Information Service-Broadcast (TIS-B), and weather and aeronautical information, called Flight Information Service-Broadcast (FIS-B). This multi-purpose data link architecture reduces significantly the operating costs. In addition, it increases the flight safety by providing traffic situational awareness, conflict detection, and alerts. An ADS-B enabled aircraft can send circumstantial information to other aircraft; this advantage allows unique situational awareness [142].

As stated before, the UAS civil aviation is still under regulation process. ADS-B Next is a part of FAA's NextGen program and will replace radars on all aircraft by 2020. ADS-B systems employ existing GPS hardware and software using real-time satellite-based internet communications.

**Conclusion:** This automatic satellite-based standard provides a great coverage range. The provided services such as situational awareness is a bonus to be employed in UASs. Even though ADS-B does not offer enough flexibility and adaptability like other standards, this standard has great potential to be used in the UAVs' communication system.

### D. *IEEE 802.11*

Wireless local area networks (WLAN) are the most popular data links used in small UAVs so far. The reasons for its popularity include easy setup, mobility support, and low cost. IEEE 802.11, commonly known as Wi-Fi, is a set of standards for implementing WLAN in 2.4, 3.6, 5 and 60 GHz bands [143]. The 802.11 and 802.11b are the oldest ones released in 1999. OFDM and direct sequence spread spectrum (DSSS) modulation are usually used in 802.11 protocols. A summary of the most popular IEEE 802.11 protocols that have potential to be utilized as data link for various UAV applications is provided in Table VI.

TABLE VI
IEEE 802.11 PROTOCOLS APPLICABLE IN UAV COMMUNICATIONS

| IEEE 802.11 protocols | Features |
|---|---|
| 802.11a | • Operates in 5.8 GHz band<br>• Data rate up to 54 Mbps depending on the type of the modulation<br>• Best tolerance against the interference compared to others |
| 802.11ac | • Operates in 5 GHz<br>• Faster than 802.11n (about 3 times)<br>• Is called the fifth (newest) generation of Wi-Fi<br>• Providing high-throughput wireless local area networks (WLANs) |
| 802.11ad | • Known as Wireless Gigabit (WiGig)<br>• Operates in 60 GHz<br>• Supports data rates up to 7 Gbps<br>Has the shortest range, which is just up to 10 m [148] |
| 802.11ah | • Also called Wi-Fi HaLow<br>• Operates at 915 MHz<br>• Suitable for low power and long-range communications such as IoT |
| 802.11b | • Operates in 2.4 GHz industrial, scientific and medical (ISM) band<br>• Data rates up to 11 Mbps, depending on the modulation type |
| 802.11g | • Operates in 2.4 GHz<br>• Data rate up to 54 Mbps |
| 802.11n | • Defines a new power save protocol known as power save multi-poll (PSMP)<br>• An improvement to the APSD protocol |
| 802.11p | • Designed for vehicular networks as Wireless Access Vehicular Environment (WAVE)<br>• Operating in 5.9 GHz frequency band for Intelligent Transportation Systems (ITS) applications |
| 802.11y-2008 | • An extension for 802.11a operations in the licensed frequency band, 3650–3700 MHz |

There are IEEE 802.11 protocols that are used specifically for network design or quality constraints. For instance, IEEE 802.11s defines a standard for wireless mesh networks describing how devices can form a WLAN multi-hop network [14]. IEEE 802.11e is a QoS scheme, defining standards for Automatic Power Save Delivery (APSD) and it considers the trade-off between energy efficiency and delivery delay for mobile devices in the data link layer.

There are several research works such as [144, 145] focusing on the design of proper data links for flying ad-hoc networks (FANETs). The proposed systems are multi-hop networks using IEEE 802.11 for the UAVs and pilot to collaborate and exchange data.

A prototype of a multi-UAV network, called UAVNet, is suggested in [146]. A framework for flying wireless mesh networks is studied. The utilized small quadcopter-based UAVs are automatically interconnected using IEEE 802.11s protocol.

A network of small UAVs, wirelessly connected through IEEE 802.11g, carrying cameras and sensors in disaster management applications have been studied in [147]. The primary goal is to have efficient cooperation among the UAVs through the network. Their proposed aerial imaging system for mission-critical situations is capable of providing services for different desired levels of detail and resolution.

There are other research studies focusing on performance measurements and tests of UASs using IEEE 802.11 as the communication network. The channel fading properties through spatial diversity, by employing multiple antennas are studied in [149]. The final scheme includes a small, low-altitude UAV within an 802.11 wireless mesh network.

Several field tests to model the UAV frequency channel are investigated in [150]. This study investigates path loss exponents for a small UAV in an 802.11a-based network using UDP (user datagram protocol). The tests on the UAV are designed to be mission-like, and the UAV goes to different waypoints, hovering around to model the behavior for a UAV needed to gather sensing information from different points. They investigate the effect of the antenna's direction on the received signal strength (RSS) and the communication throughput of the UAV's data link flying at different altitudes. In research works [151, 152] the main focus is to improve the communication links between the UAVs and UAVs to the pilot by conducting several tests. They consider throughput and radio transmission range as performance metrics and use 802.11n and 802.11ac in the infrastructure and mesh structure among the UAVs. Their results prove that 802.11n and 802.11ac provide high throughput along with high data rates.

**Conclusion:** IEEE 802.11 standards have been used worldwide, since they can function in a wide range of frequencies. This huge deployment has led to the maturity of these standards, which is an advantage of this standard [153]. Other benefits of this standard can be mentioned as they are easy to setup with low cost, and they support adequate level of mobility. However, considering its disadvantages from the aeronautical standardization point of view, they are considered as short-range wireless LANs. Also, they have not been tested officially for aeronautical use by any official standardization body yet. Another downside of these data links is the high level of interference in the license-exempt bands that might cause problems for mission-critical UAVs. Moreover, these standards were not initially designed for aeronautical or aviation purposes, although they have been widely employed for unmanned aerial applications.

## *E. VDL Mode 2*

The VHF Digital Link (VDL) is used in aircraft as a data link to communicate with the ground station. It was defined by ICAO Aeronautical Mobile Communications Panel (AMCP) in the 1990s. They offered VDL technology since the data rate of 2.4 kbps of analog VHF radios was not sufficient anymore for aviation communication, and the analog radios used for voice data could not support digital communications. The set of proposed digital data links operating in VHF band was called by the general name of VDL.

VDL consists of four operational modes. Mode 1 was the first version using analog radios. Mode 3 was an unsuccessful plan trying to provide aircraft digital communication for both data and voice. Mode 4 was initially intended as a physical layer for ADS-B transmissions, but no longer exists. Hence, mode 2 is the only VDL data link that is still in operation [154]. Therefore, the terms "VDL" or "VDL2" are generally used as abbreviations for "VDL mode 2," as done in this paper also.

VDL operates at 117.975 and 137 MHz of the aeronautical VHF band. It uses differential 8 phase shift keying (D8PSK) modulation, with a data rate up to 31.5 kbps. The VDL data frame is shown in Fig. 7.

Data transmission begins with a training sequence for demodulation. Training sequence consists of five parts: 5 symbols for transmitter power stabilization; 16 symbols for synchronization; one symbol is reserved symbol; one 17-bit word to indicate the transmission length; and 5 bits for header forward error correction (FEC). An Aviation VHF Link Control (AVLC) data frame follows the training sequence [155, 156]. The AVLC frame format is adapted from high-level data link control (HDLC) protocol, and it contains the source and the destination addresses, the payload data and a frame check sequence (FCS), all added to the end of the frame [157].

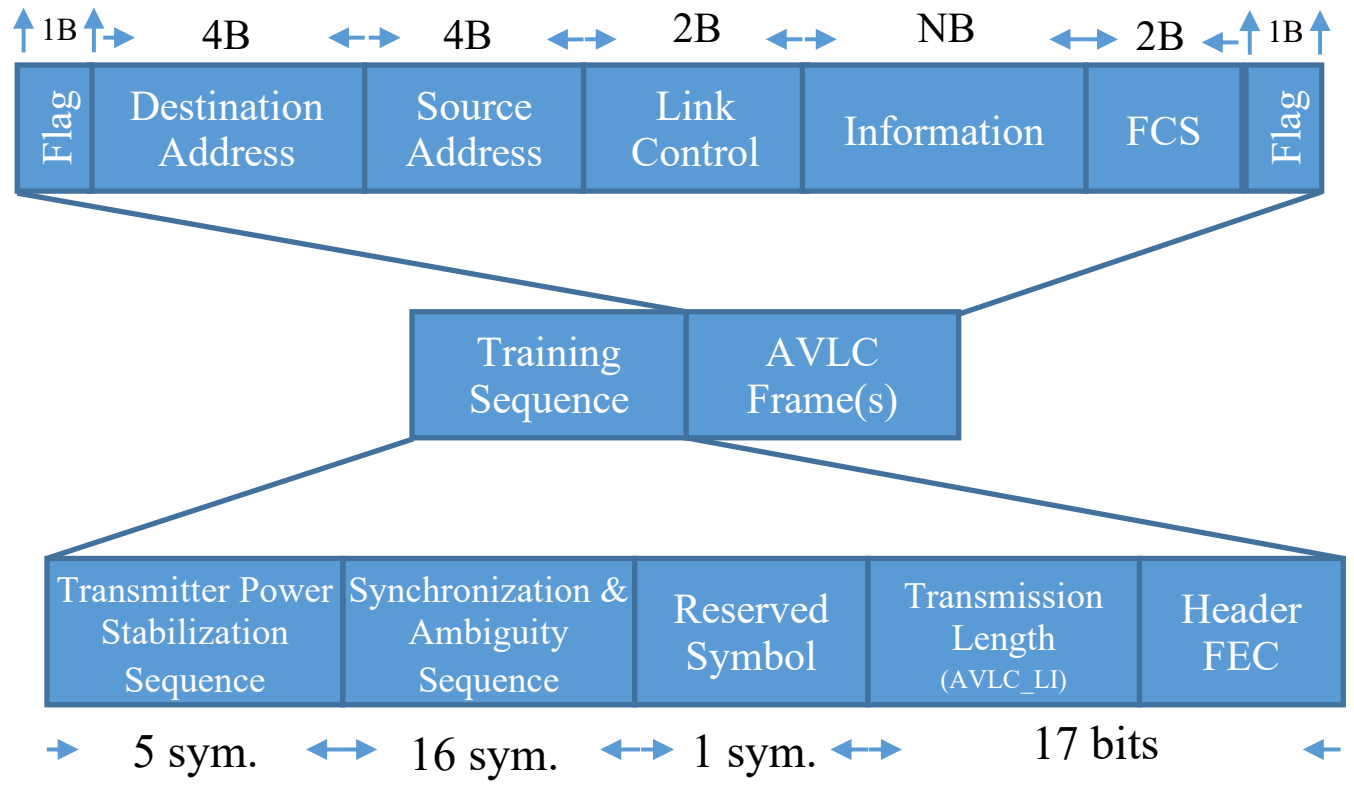

Fig. 7. VDL Mode 2 Data Frame. Each symbol (sym.) is three bits. B is short for Byte.

VDL2 has been widely used for Aircraft Communications Addressing and Reporting System (ACARS) messages as a digital data link for aircraft to exchange short length data. It is known as ACARS-Over-AVLC (AOA) since the AVLC is the data link layer of the VDL2 standard [158]. VDL2 is also applicable as the data link for Aeronautical Telecommunications Network (ATN) using the same VHF ground and aircraft radios. Even though, ATN standard is slightly different from ACARS [159].

**Conclusion:** VDL2 is a data link that has been standardized for aviation. It has been already employed in UASs, and its broad coverage area makes it practical for most of the applications. It can support a high level of mobility and robustness, which are critical features for UAV communications. However, the lack of flexibility and adaptability is a challenge restricting the VDL popularity and applicability in a wide range of UAV applications.

## *F. CPDLC*

Controller pilot data link communication (CPDLC) is a message-based service for manned aviation communications between the pilot and the ATC. CPDLC usually uses satellite communication and VDL2 data links for ATC communications. InmarSAT was trying to get the certification to be allowed to tunnel CPDLC over their satellite internet protocol links [160]. Until 2016, Iridium satellite was the only network authorized to their communication link with CPDLC.

The message elements provided by CPDLC are categorized as "clearance," "information" or "request," which has the same phraseologies used in the radiotelephone. The ATC uses CPDLC via a terminal to issue clearances, to exchange information, and to answer the messages such as required instructions, advisories, and emergency guidance. The pilot can reply to the messages, to request clearances, to exchange information, and to announce or call in an emergency. The communication happens by selecting predefined phrases (e.g., EMERGENCY, REPORT, CLEARANCE, REQUEST, LOG, WHEN CAN WE, etc.) Further, both the pilot and the ATC can also exchange free text messages which do not follow the pre-defined formats [161].

CPDLC satisfies the communication requirements to meet the CNS demands for the future global aviation [162]. CPDLC is secure; thus, it is popular for communicating confidential and critical aviation information [163].

Safety analysis of employing CPDLC for unmanned systems has been studied in [6]. A faulty communication model has been tried out to examine the risks associated with the interference in CPDLC communications. However, as they assert the results, it is practical to use CPDLC in the UAS communication, but some adaptation is necessary to guarantee the message's integrity.

**Conclusion:** CPDLC can assure the required safety for the data link used in the UASs. Features such as robustness, easy to employ, and efficiency make it useful for sensitive applications where failure is not accepted. To be used as the general standard aviation platform for UAVs, modifications must be applied. Using satellite or terrestrial data links along with CPDLC will affect the coverage range and the financial cost of the system.

### *G. SWIM*

As mentioned before, FAA is focusing on its future global plan for aviation standardization, NextGen, planning to roll out by 2025. System Wide Information Management (SWIM) is the main part of this plan. FAA is focusing on using SWIM to provide a secure platform for cooperation among the national and international aviation organizations. SWIM concept was first initiated by ICAO to improve the data access for all the elements in the network [164]. It is supposed to turn the NAS into a network-enabled system and use the exchanged data from the ATM networks to improve the traffic management, safety, and situational awareness [165].

Enterprise Service Bus (ESB) is one of the leading parts of the SWIM architecture. The primary role of the ESB is to provide a general middleware for the basic communication between different service users. ESB is also expected to provide higher level functions including content-based routing, data monitoring for fault management and security functions, password management, authentication, and authorization. Further, secure gateways to communicate with all non-NAS users are needed [166].

SWIM must be able to provide communication links between the NAS and non-NAS users. These services will follow the “publish/subscribe” or “request/reply” messaging patterns. They all will use one of the four data exchange standards; Aircraft Information Exchange Model (AIXM); Weather Information Exchange Model (WXXM); Flight Information Exchange Model (FIXM), or the proposed ICAO ATM Information Reference Model (AIRM) [167].

The aviation data link of this platform is called Aircraft Access to SWIM (AAtS) that manages the communication between the aircraft and the ground station through cellular or satellite communication networks. However, the functionality of this data link is limited to just advising services and not actually controlling the aircraft. Hence, this data link is not used for command and control and is only for planning and awareness. Further, an intermediary service will be implemented in SWIM called Data Management Service (DMS) providing a data link for piloting the aircraft [168].

**Conclusion:** since SWIM has not been implemented completely, we cannot judge its performance or suitability for UAVs yet. However, the anticipated features by FAA sound very promising to pave the way of integrating all manned and unmanned vehicles into the same airspace. This service is supposed to provide a safe platform for communications and offering a higher level of situational awareness.

### *H. Preliminary Study*

We have done a preliminary study [169]. In this work, we have investigated the limitations that AeroMACS imposes on the UAV communications in theory, when it comes to integrating the UAVs into the NAS, using AeroMACS.

Signal to inter carrier interference (ICI) ratio and channel coherence time are the two channel modeling properties that we utilized to find the tolerance threshold of the UAV’s speed. ICI is simply the phenomenon of signal’s carriers overlapping. This occurrence is caused by Doppler shift and deforms the signal’s shape at the receiver.

Coherence time is another channel property. It shows the time duration that we can assume the channel impulse response will not change. Therefore, calculating the channel model estimation is not necessary during this time. The larger the coherence time, the better, since the system’s complexity will be less due to no need to run channel estimation very often, which can be very sophisticated.

We calculated these two important parameters, based on the specific properties of AeroMACS. As shown in Fig. 8, as the aircraft’s speed increases, the signal’s power will be lost in the ICI power. Also, the channel coherence time decreases by about three-fold, which is actually a lot in this application. As explained before, the main reason of performance degradation is the Doppler shift caused by the aircraft mobility. We theoretically prove that the maximum tolerable speed of UAV using AeroMACS data link is around 35.9 m/s (129.25 km/h). This actually needs to be improved through the AeroMACS standard, as if it becomes an essential standard in the unmanned area. For more detail, we refer the readers to the paper [169].

### *I. Summary*

In this section, we focused on several aviation standards that have potentials to be used for UAVs communications. Data links for unmanned aviation are still being studied in standards bodies. Due to its unique requirements, none of the current aviation standards can be considered as a standalone standard without further modifications. These standards were primarily developed for manned aircraft and have limitations when it comes to unmanned flights.

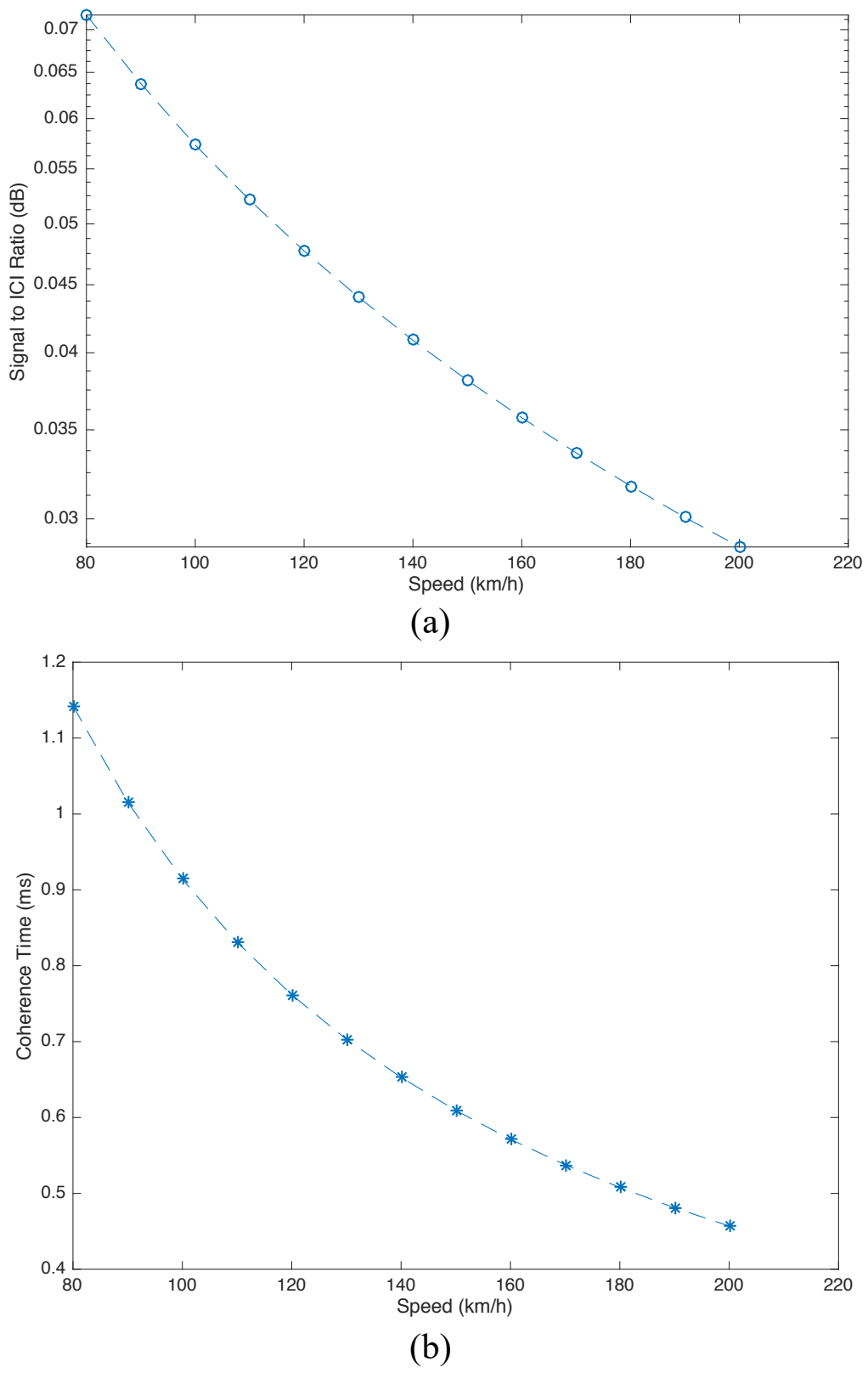

Fig. 8. (a) Signal to ICI Ratio. (b) Channel coherence time

Aviation standards are designed to support a high level of mobility. Hence, these standards do not have Doppler shift and coherency problems. However, since the data link requirements are completely dependent on the type of the UAV's mission, adaptability and flexibility are important. UAV's applications range from hobbies to mission-critical tasks, such as rescue missions, and, therefore, require very different features and demands. Table VII summarizes the main advantages and disadvantages of each standard, along with their operating frequencies. Since CPDLC and SWIM are mostly used as secured tunneling platforms for other data links for the aircraft flights, there are excluded from the table.

TABLE VII
COMPARISION OF DIFFERENT AVAILABLE AVIATION STANDARDS

| Standards | Operating Frequency | Advantages | Disadvantages |
|---|---|---|---|
| AeroMACS | 5091-5150 MHz | • Flexibility<br>• Adaptability | • Lower mobility |
| L-DACS | 960 -164 MHz | • High mobility<br>• High coverage | • No flexibility<br>• Low coverage |
| ADS-B | 978 & 1090 MHz | • High mobility<br>• Situational awareness<br>• High coverage | • No flexibility<br>• Relatively expensive |
| IEEE 802.11 | 915 MHz, 2.4 & 5.8 GHz | • High data rate<br>• Maturity | • Lower mobility |
| VDL Mode2 | 117.975 & 137 MHz | • High mobility<br>• Robustness<br>• High coverage | • No flexibility |

## VII. STANDARDIZATION ORGANIZATIONS

Different international standard organizations are trying to establish their own standard framework for UAS data links. The European Commission started a 2-year project to test and propose a secure architecture for a UAV-based surveillance system. In their project, there were several work packages (WPs). One of them, WP3, was dedicated to looking for an effective data link for UAS communication. One of the main tasks of this WP (WP3.3) was dedicated to analyzing all available LOS and BLOS data links, to pick the best candidates, and to improve them even further for better integrity and availability. However, their final results indicated the lack of any comprehensive standard communication service or specific frequency bands specialized for UAS [170].

In this section, we highlight some of the standard bodies that are actively working on developing new standards for UASs. The primary focus of these standard bodies is to ensure safe flights for all aircraft given that the unmanned vehicles are officially a part of the NAS and will affect the safety at a large scale. The organizations that we discuss are RTCA, ASTM, EUROCAE, and 3GPP.

### *A. RTCA*

Radio Technical Commission for Aeronautics (RTCA), founded in 1935 in the United States, provides technical guidance for industry and governmental aviation applications voluntarily. The performance standards developed by RTCA form the basis of FAA regulatory requirements to ensure safety for the air transportation today [171]. This organization has more than 200 committees serving as federal advisory.

RTCA's unmanned aviation sector started in 2013. It is known as the Special Committee (SC)-228 and consists of two working groups, Detect and Avoid (DAA) working group, WG-1; and Command and Control (C2) working group, WG-2. They develop terrestrial data links operating in L band and C band.

The current DAA WG-1 has focused on several concepts, including modeling and simulating the humans' effect in the aviation loop, UAV operations under IFR, and transition through different airspace classes. It is planned to be extended to cover operations in all airspace classes. This committee cooperates with RTCA SC-147, the Traffic Alert & Collision Avoidance System (TCAS) sector, to provide comprehensive standards.

C2 WG-2 aims to provide minimum operational performance standards (MOPS) for dynamic resource allocation, security and safety, levels of automation, and human factors. Also, they investigate the possibility of simultaneous operations and hardware integration of L band and C band to enhance efficiency and system performance and ensure that C2 MOPS meets DAA needs [172, 173]. Phase 1 of WG-2 focused on large aircraft based on operational SWaP. RTCA developed several tests on verification and validation (V&V) of initial MOPS for civilian UASs.

The practical aspects of these MOPS on terrestrial radio waveform concerning SWaP limitations of UAVs have been studied in [174]. They offer several suggestions on the physical layer data structure to improve the performance of the system.

At the end of 2015, RTCA made an important announcement about standardizing the required safety mechanisms regarding collision avoidance for large UAVs [175]. The standard is related to the communications between the remote pilot and the UAV. More specifically, the scope of this standard is about the required MOPS for "detect and avoid" through command and control links. This milestone is considered as a big step toward standardizing the UAV flights.

Meanwhile, smaller UAVs, which are predicted to be deployed in larger numbers than considered previously, also need to be able to operate BLOS missions, while they have significant SWaP limitations. SATCOM C2 link for small UASs is in the second phase of the WG-2 and is expected to be finalized by 2020 [176]. Regarding the SATCOM data link properties, Phase 2 activities are mainly trying to extend the point-to-point communication model of the first phase to enable the UAVs with the BLOS operations and applications. Multiple bands, including selected Ku and Ka sub-bands, and possibly C band allocations are planned to be used [177]. The SATCOM UAS C2 data link MOPS, provided by RTCA, is dedicated to the performance requirements of satellite-based UAS data links. It also includes recommendations for a V&V test and performance analysis.

There is another special committee, SC-223 from RTCA that is developing standards on AeroMACS. The developed AeroMACS MOPS and standards in this committee are supposed to be applied in the future SWIM platform as well.

RTCA, like all other aviation standard developers, should address future requirements driven by the increasing interest in the UAS market and the effect on command and control performance in addition to implementation limitations.

### *B. ASTM*

American Society for Testing Materials (ASTM) was originally a United States standard developing industry association that was established in 1898. It is now called ASTM International, and it works on various products and services related to a wide range of safety aspects including aviation and aerospace standardizations [178].

The committee F38 under ASTM International, dedicated to standardizing civilian unmanned air vehicle systems, started in 2011. They work on safety concerns related to design, performance analysis, practical examinations and tests [179].

In 2015, the organization started a program entitled "Standard Practices for Unmanned Aircraft System Airworthiness." As mentioned in [180], this program works on classifying existing regulations and standards about the design, manufacturing, and maintenance of UAVs. The primary objective of this standard body is to provide certificates for aircraft's safe flight assurance to the UAS designers and manufacturers based on governmental aviation requirements.

Some of the other recent standards developed for UASs by ASTM are "Standard Practice for Unmanned Aircraft System (UAS) Visual Range Flight Operations," which were requested by the FAA. This standard is on LOS operations of UAVs helping secure their integration into the national airspace.

ASTM has divided the UAS standard practices into two groups: "standard airworthiness certificate for large UAS" and "standard airworthiness certification for light UAS," also called Light-Sport Aircraft (LSA). The UAS must meet several requirements defined by FAA to be categorized as an LSA. In particular, the weight of the UAV must be less than 600 kg. Further, the UAV manufacturers should provide the necessary documents showing they are following the required standards. However, the only criteria considered by ASTM is that the maximum take-off weight of the UAV must be at most 600 kg. The ASTM is currently working on defining standard frameworks for UAVs under 25 kg [181].

### *C. EUROCAE*

The European Organization for Civil Aviation Equipment (EUROCAE) is the only European standardizing body exclusively devoted to the development of aviation technical standards [182]. EUROCAE is similar to RTCA in the United States. This organization has several technical Working Groups (WGs).

EUROCAE WG-73 is related to UAS standards and was created in early 2007. This group focuses on unmanned flights in airspace classes A, B, and C in the context of European ATM. One of the sub-groups (SG) of WG-73, which was working on light Remotely Piloted Aircraft Systems (RPAS) operations, got separated from the group and was established as a new group, EUROCAE WG-93 in 2012. This group works on LOS RPAS weighing less than 25 kg. They classify these aircraft into 4 categories; Harmless: less than 250 g; A0: less than 1 kg, A1: Less than 4 kg, A2: Less than 25 kg [5].

EUROCAE WG-73 is equivalent to RTCA SC-228. Currently, WG-73 consists of four sub-groups: SG1 related to UAS operations; SG2 related to UAS airworthiness; SG3 related to command, control, communications, spectrum and security; and SG4 related to UAVs under 150 kg.

In November 2010, WG-73 published a paper entitled "A Concept for UAS Airworthiness Certification and Operational Approval" (EUROCAE-ER-004). This paper discusses the European Aviation Safety Agency (EASA) regulation regarding the UAVs. Such regulations include UAS airworthiness specifications and general requirements and considerations for any UAS manufacturer or designer to be qualified for EASA certifications [183].

### *D. 3GPP*

Seven telecommunications standard organizations, known as "Organizational Partners," form the 3rd Generation Partnership Project (3GPP). Their objective is to specify and standardize the telecommunication technologies, such as LTE, EDGE, GPRS, etc. It was initiated in the United States in 2004, and they partnered with organizations from Asia (including Korea, Japan, China, and India) and Europe [184]. The 3GPP started by developing standards for 3G mobile phone systems. They have continued to develop new releases for 4G and 5G. The newest release, which is release number 15, along with the release number 14, is focused on the standardization of the 5G mobile network.

The 3GPP LTE cellular network has been suggested as a communication system for UAVs in recent years, as we discussed in Section V of this paper. The main advantage of using LTE in UASs is in providing the required ATC and ATM services for the system. This ensures safety and efficiency of shared airspace among manned and unmanned vehicles [185].

Major companies in 3GPP, such as Qualcomm, have been working on further optimizing LTE for UAVs. They have announced a set of testing results on UASs using 4G LTE network as the communication data link at the Qualcomm UAS Flight Center in San Diego. They ran over a thousand trial flights measuring the key performance indicators (KPIs) including coverage, signal strength, throughput, latency, and mobility. They implemented different scenarios through the LTE network and examined the cellular system performance in networks serving low-altitude (122 m AGL and below) UAVs. The results showed that 4G LTE networks perform very well for drones operating up to 122 m AGL [186].

Unlike other standard-related bodies, 3GPP has not dedicated any working or sub groups to UAVs and their issues. However, since cellular technology is very promising for UAVs, there is a need to establish a group in 3GPP to consider the standards related to UAV's data links and their constraints.

TABLE VIII
COMPARISON OF DIFFERENT STANDARDS ORGANIZATIONS

| Organization | Year of Activation in Unmanned Area | Groups dedicated to UAV Flights | Main Focus |
|---|---|---|---|
| RTCA | 2013 | SC-228:<br>• WG-1, related to DAA issues<br>• WG-2, related to C2 issues<br>SC-223, developing MOPS and standards for AeroMACS | • Terrestrial and satellite data links in C & L bands<br>• Improving the MOPS for safe UAV flights |
| ASTM | 2011 | F38, dedicated to standardizing civilian UASs | • Certificates on safety aspects to ensure safe UAV flights<br>• Mainly focusing on sUAVs weighing less than 25 kg |
| EUROCAE | 2007 | WG-93, related to light UAVs (under 25 kg)<br>WG-73:<br>• SG1, related to UAS operations<br>• SG2, related to UAS airworthiness<br>• SG3, related to C2 communications, spectrum, and security<br>• SG4, related to UAVs under 150 kg | • Standardizing unmanned flights in different airspace classes<br>• Airworthiness specifications and regulations |

### E. Summary

In this section, we described different organizations related to standardizing aviation platforms that are actively working in unmanned areas as well. However, to come up with a comprehensive and international standard for UAV data links, all these organizations must cooperate. Table VIII summaries specific features related to each of these organizations. The year of their activation in the UAV area is also included. We have excluded 3GPP since they do not have any dedicated sub group for unmanned applications.

## VIII. FUTURE DIRECTIONS

In this section, we discuss some of the challenges that are associated with UAV data links. Further, for each item, we talk briefly about the primitive solutions that may be tried to solve the problems. Although each of these proposals can be argued for and against, we are not proposing these as final solutions. More research is required on each of these issues. The main goal here is to ease the way of standardization process for this sector of aviation. When it comes to solving a challenge, spectrum, SwaP limitations, signal propagation, and routing must be considered.

***SATCOM****:* The *high latency* (compared to other data link candidates) is the main challenge. Considering the long distance between the satellite antenna and the UAVs close to the earth surface, *real-time communication* is not practical. Moreover, *large and heavy antennas* that must be mounted on the UAV for sending and receiving are another obstacle in this area. *High level of attenuation and signal loss* is another challenge in using satellite as the main communication network. Moreover, the *signal interference* that SATCOM services face with some terrestrial services is another issue. To provide a practical standard spectrum dedicated to UAVs to use SATCOM all these restrictions must meet reasonable solutions first.

One of the primitive ***solutions*** can be using *multi-hop communications* to break down the link between the UAV and the satellite. Taking advantage of the *edge computing* technology and implementing a join design of edge and satellite will surely solve the latency issue. Another solution is employing effective *automation* to pre-program the whole mission. This will eliminate the need for low-latency data links. Defining and dedicating a proper bandwidth for UAV aviation would also solve the issue of interference. These solutions will also decrease the antenna size and the power consumption.

***Cellular:*** The biggest issue for cellular communication is that they are *not designed for aviation*. Hence, UAV requirements (or any other type of aircraft) have not be considered in the design. *Congested bands* that get even worse in crowded areas is another challenge. *Short range coverage* and the need for handover techniques bring up more issues. More research on *measuring the performance* in the congested area and handover deficiencies are required.

The primitive ***solution*** for this matter is *to define a separate band for UAVs*. This not only helps the congestion challenge but also avoids interferences with other non-aeronautical users. Also, since the number of users (i.e., the UAVs) at each coverage zone using that band would be low, the dedicated band can support much longer areas. On the other hand, to solve the issue of short-range coverage of cellular towers and the need for an optimized handover, special on-board processing can be designed for the UAVs. The designed processor can utilize the provided Internet by cellular network and find the optimized rout and help the handover procedure.

**Current Aviation Standards:** There are different drawbacks of the available standards, such as *lack of adaptability and limited bandwidth*. Other limitations such as *short-range coverage* and *high level of interference* are also challenging, for example in Wi-Fi communication link. Since most of the discussed standards are a part of the NextGen plan, these limitations must be solved in the near future.

Regarding these issues, the foremost ***solution*** is to design a data link standard with flexibility feature offering different levels of QoS, data rate, latency, and robustness. This adaptability will satisfy a wide range of UAV applications requirements. While integrating the flexibility and adaptability features in the data link, various applications of UAVs should be considered. For instance, for imaging applications, a higher bandwidth must be dedicated to the payload downlink, while in delivery applications, control link plays the fundamental role and must occupy most of the band. A specially dedicated data link standard that operates in a protected spectrum will not have the problem of short range or interference. The operating frequency cannot be too high due to the signal propagation issues and cannot be too low due to the congested spectrums. Hence, a balance study must be conducted to consider the trade-offs.

## IX. SUMMARY

In this paper, we provided a comprehensive survey of data link technologies that can be employed for civilian UASs for various applications. First, we discussed the basic requirements that all civilian data links should provide, and general properties of unmanned aerial data links. After that, we studied the satellite data links, allowing UAS to have BLOS operations and extending their coverage area. BLOS capability helps the system detect any upcoming collision early and significantly improves the safety of the system. Currently, there are many active companies in satellite communication, and we discussed some of the most popular ones, their latest product, and their service features. It is critical to notice that the best choice of the data link is highly dependent on the application for which the system will be used.

Following that, we discussed cellular data links. We mentioned their advantages and disadvantages over other techniques. By rolling out the ultra-fast and high-performance 5G in the near future, the cellular technology could be the most powerful candidate for the unmanned systems. Employing the cellular network as the data link for UASs would boost the functionality of autonomous flight in mission-critical services. This would help save lots of human and animal lives in natural disasters or wildfires. The new cellular generation would impact other areas of the economy. Since, the unmanned vehicles will be the main segment in goods delivery, hobbyist, and commercial applications. The cellular data links, regardless of the potential high expenses, would supply the system with QoS requirements and high level of performance. Moreover, cellular services provide the system with high availability and broader coverage.

After that, we discussed related standards such as AeroMACS, L-DACS, ADS-B, IEEE 802.11, VDL Mode 2, CPDLC, and SWIM. These standards are currently being discussed as the potential for unmanned data links. We mentioned their structure and applications. These standards need to be improved and adapted further to ensure that all co-existing unmanned and manned flights in the NAS would not be harmful to the environment or other aircraft. Without providing enough level of safety in the standards for the increasing number of unmanned aircraft in the future, the chaos caused by all unorganized flights would be catastrophic and will create a hazard to the environment and everyone. We also brought up the results of our preliminary study to show the theoretical limitations that AeroMACS, as an example standard, puts on the UAV speed limit.

Then, we presented four leading active standard bodies in unmanned aviation, RTCA, ASTM International, EUROCAE, and 3GPP. Unmanned aviation has gained great interest, and almost all standard organizations in aviation have dedicated separate working groups to this area. However, unmanned aviation still suffers from a lack of maturity of a comprehensive set of standards that would guarantee the required capacity and safety for the future growth. Hence, this is where the standard bodies have focused on recently, and significant global cooperation among all the continental organizations is required.

Finally, based on the concepts that were studied during this paper, future challenges associated with this area of research were highlighted. Some primitive solutions to be considered to solve each problem have been suggested as well.

## APPENDIX

### A. *Table of Abbreviations*

The acronyms and their definitions used throughout this survey paper are provided below.

TABLE IX
LIST OF ABBREVIATIONS

| *Abbreviation* | *Explanation* |
|---|---|
| 1090ES | 1090 MHz Extended Squitter |
| 3G | Third Generation |
| 3GPP | 3rd Generation Partnership Project |
| 4G | Fourth Generation |
| 5G | Fifth Generation |
| AAC | Airline Administrative Communications |
| AAtS | Aircraft Access to SWIM |
| ABS | Anti-lock Braking System |
| ACARS | Aircraft Communications Addressing and Reporting System |
| ADS-B | Automatic Dependent Surveillance-Broadcast |
| ADSL | Asymmetric Digital Subscriber Line |
| AeroMACS | Aeronautical Mobile Airport Communications System |
| AeroWAN | Aeronautical Wide-Area Network |
| AGL | Above Ground Level |
| AIA | Aerospace Industries Association |
| AIRM | ATM Information Reference Model |
| AIXM | Aircraft Information Exchange Model |
| ALAS | ADS-B Link Augmentation System |
| AM(R)S | Aeronautical Mobile (Route) Service |
| AMACS | All-purpose Multichannel Aviation Communication System standard |
| AMCP | Aeronautical Mobile Communications Panel |
| AMS(R)S | Aeronautical Mobile Satellite (Route) Service |
| AOA | ACARS-Over-AVLC |
| AOC | Airline Operational Communications |
| APCO | Association of Public Safety Communications Officials |
| APSD | Automatic Power Save Delivery |
| ASTM | American Society for Testing Materials |
| ATC | Air Traffic Control |
| ATM | Air Traffic Management |
| ATN | Aeronautical Telecommunications Network |
| ATO | Air Traffic Organization |
| AVLC | Aviation VHF Link Control |
| B-AMC | Broadband Aeronautical Multicarrier Communication |
| BER | Bit Error Rate |
| BGAN | Broadband Global Area Network |
| BLOS | Beyond the Line of Sight |
| BS | Base Station |
| C2 | Command and Control |
| CDMA | Code Division Multiple Access |
| CIR | Channel Impulse Response |
| CNPC | Control and Non-Payload Communication |
| CNS | Communications, Navigation, and Surveillance |
| COA | Certificates of Authorization or Waiver |
| COC | Clear of Clouds |
| CoMP | Coordinated Multi-Point |
| CPDLC | Controller Pilot Data Link Communication |
| CS | Coding Scheme |
| D2D | Device to Device |
| D8PSK | Differential Encoded 8 Phase Shift Keying |
| DAA | Detect and Avoid |
| DMS | Data Management Service |
| DSSS | Direct-Sequence Spread Spectrum |
| EASA | European Aviation Safety Agency |
| EDGE | Enhanced Data rates for Global Evolution |
| EMI | Electromagnetic Interference |
| ESB | Enterprise Service Bus |
| EUROCAE | The European Organization for Civil Aviation Equipment |
| FAA | Federal Aviation Administration |
| FANET | Flying Ad hoc Network |
| FBMC | Filter-Bank Multi-Carrier |
| FCC | Federal Communications Commission |
| FCI | Future Communication Infrastructure |
| FCS | Future Communication System |
| FCS | Frame Check Sequence |
| FEC | Forward Error Correction |
| FIS-B | Flight Information Service-Broadcast |
| FIXM | Flight Information Exchange Model |
| FSS | Fixed Satellite Service |
| GEO | Geosynchronous Earth Orbiting |
| GMSK | Gaussian Minimum Shift Keying |
| GNSS | Global Navigation Satellite System |
| GPRS | General Packet Radio Service |
| GS | Ground Station |
| GSM | Global System for Mobile communication |
| GX | Global Xpress |
| HD | High Definition |
| HDLC | High-Level Data Link Control |
| HTS | High-Throughput Satellites |
| ICAO | International Civil Aviation Organization |
| ICI | Inter Carrier Interreference |
| IFR | Instrument Flight Rules |
| IoT | Internet of Things |
| ISDN | Integrated Services Digital Network |
| ISM | Industrial, Scientific and Medical |
| ITS | Intelligent Transportation Systems |
| ITU | International Telecommunication Union |
| JAA | Joint Aviation Authorities |
| KA | Keep-Alive |
| KPI | Key Performance Indicator |
| L-DACS | L-band Digital Aeronautical Communication Systems |
| LDL | L-band Data Link |
| LEO | Low Earth Orbiting |
| LFS | Losses in Free Space |
| LOS | Line of Sight |
| LSA | Light-Sport Aircraft |
| LTE | Long-Term Evolution |
| M2M | Machine to Machine |
| MAC | Medium Access Control |
| MEO | Medium Earth Orbit |
| MIMO | Multiple Input Multiple Output |
| MOPS | Minimum Operational Performance Standards |
| MPR | Multi-Packet Reception |
| MSL | Mean Sea Level |
| NAS | National Airspace System |
| NextGen | Next Generation |
| NLOS | Non-Line of Sight |
| OFDM | Orthogonal Frequency-Division Multiplexing |
| OG2 | Orbcomm Generation 2 |
| PCS | Personal Communication Services |
| PPP | Point-to-Point Protocol |
| PSMP | Power Save Multi-Poll |
| QoS | Quality of Service |
| RF | Radio Frequency |
| RPAS | Remotely Piloted Aircraft Systems |
| RSS | Received Signal Strength |
| RTCA | Radio Technical Commission for Aeronautics |
| RTCA | Radio Technical Commission for Aeronautics |
| RTL | Round-Trip Latency |
| SATCOM | Satellite Communication |
| SC | Special Committee |
| SG | Sub-Groups |
| SINR | Signal to Interference plus Noise Ratio |
| SNR | Signal to Noise Ratio |
| STC | Supplemental Type Certificate |
| SVFR | Special Visual Flight Rules |
| SWaP | Size, Weight, and Power |
| TCAS | Traffic Alert & Collision Avoidance System |
| TCP | Transmission Control Protocol |
| TDD | Time-Division Duplex |
| TIA-902 | Telecommunications Industry Association Standard 902 |
| TIS-B | Traffic Information Service-Broadcast |
| TOA | Time of Arrival |
| UA | Unmanned Aircraft |
| UAS | Unmanned Aircraft System |
| UAT | Universal Access Transceiver |
| UAV | Unmanned Aircraft Vehicle |
| UDP | User Datagram Protocol |
| UMTS | Universal Mobile Telecommunications System |
| USV | Unmanned Satellite Vehicle |
| V&V | Verification and Validation |
| VDL | VHF Data Link |
| VFR | Visual Flight Rules |
| VMC | Visual Meteorological Conditions |
| W-CDMA | Wideband Code Division Multiple Access |
| WAVE | Wireless Access Vehicular Environment |
| WG | Working Group |
| WiGig | Wireless Gigabit |
| WLAN | Wireless LAN |
| WP | Work Package |
| WRC | World Radiocommunication Conference |
| WXXM | Weather Information Exchange Model |